\documentclass[prl,reprint,superscriptaddress,floatfix]{revtex4-2}

\usepackage{pdfpages} 
\usepackage{pgffor}
 \usepackage{comment}
 \usepackage{amsmath}
 \usepackage{amsfonts}
 \usepackage{graphicx}
 \usepackage{comment}
 \usepackage{xcolor}
 \usepackage{epstopdf}
 \usepackage{newtxtext,newtxmath}
\usepackage{dcolumn}
\usepackage{physics}
 \usepackage{hyperref} 
 \usepackage{epstopdf}

 \hypersetup{
    colorlinks=true,       
    linkcolor=blue,          
    citecolor=blue,        
    filecolor=magenta,      
    urlcolor=blue           
}

\renewcommand{\epsilon}{\varepsilon}

\allowdisplaybreaks

\newcolumntype{d}[1]{D{.}{.}{#1}}

\let\originalleft\left
\let\originalright\right
\renewcommand{\left}{\mathopen{}\mathclose\bgroup\originalleft}
\renewcommand{\right}{\aftergroup\egroup\originalright}

\makeatletter
\AtBeginDocument{\let\LS@rot\@undefined}
\makeatother

\begin{document}

\frenchspacing

\title{Many-body Theory Calculations of Positron Binding to Halogenated Hydrocarbons}
\author{J. P.~Cassidy}
\email{jcassidy18@qub.ac.uk}
\affiliation{
School of Mathematics and Physics, Queen's University Belfast, University Road, Belfast BT7 1NN, United Kingdom}
\author{J. Hofierka}
\affiliation{
School of Mathematics and Physics, Queen's University Belfast, University Road, Belfast BT7 1NN, United Kingdom}
\author{B. Cunningham}
\affiliation{
School of Mathematics and Physics, Queen's University Belfast, University Road, Belfast BT7 1NN, United Kingdom}
\author{C. M. Rawlins}
\affiliation{
School of Mathematics and Physics, Queen's University Belfast, University Road, Belfast BT7 1NN, United Kingdom}
\author{C.~H.~Patterson}
\affiliation{
School of Physics, Trinity College Dublin, Dublin 2, Ireland
}
\author{D.~G. Green}
\email{d.green@qub.ac.uk}
\affiliation{
School of Mathematics and Physics, Queen's University Belfast, University Road, Belfast BT7 1NN, United Kingdom}

\date{\today}


\begin{abstract}
Positron binding energies in halogenated hydrocarbons are calculated \emph{ab initio} using many-body theory.
For chlorinated molecules, including planars for which the interaction is highly anisotropic, very good to excellent agreement with experiment and recent DFT-based model-potential calculations 
is found. 
Predictions for fluorinated and brominated molecules are presented. 
The comparative effect of fluorination, chlorination and bromination is elucidated by identifying 
trends within molecular families including dihaloethylenes and halomethanes based on global molecular properties (dipole moment, polarizability, ionization energy). 
It is shown that relative to brominated and chlorinated molecules, fluorinated molecules generate a less attractive positron-molecule potential due to larger ionization energies and smaller density of 
molecular orbitals close to the HOMO,
resulting in very weak, or in most cases loss of, positron binding. 
Overall, however, it is shown that the global molecular properties are not universal predictors of binding energies, exemplified by consideration of CH$_3$Cl vs.~\emph{cis.}-C$_2$H$_2$F$_2$: 
despite the latter having a larger dipole moment, lower ionization energy and similar polarizability its binding energy is significantly smaller (25 meV vs.~3 meV, respectively), owing to the important contribution of multiple molecular orbitals to, and the anisotropy of, the positron-molecule correlation potential. 
\end{abstract}

\maketitle

Trap-based positron beams have enabled resonant-annihilation-based measurements of positron binding energies for around 90 molecules 
\cite{Gilbert02,RevModPhys.82.2557,Danielson09,Danielson10,Danielson12,Danielson12a,Swann2021effect,Danielson_2021,Ghosh:2022,Danielson:2022}. 
Whilst the corresponding theory of positron capture into vibrational Feshbach resonances is well established \cite{PhysRevA.61.022720, Gribakin2001,RevModPhys.82.2557}, accurate calculations of positron binding energies have  been realised only relatively recently (see e.g., \cite{Tachikawa11,Tachikawa14,APMO2014,Swann:2018,Swann:2019,Swann:2020,Swann2021effect,molbind_dft4,MLbind,Jaro22}).
Attempts have been made to relate the observed binding energies to the global molecular properties including the dipole moment $\mu$, isotropic polarizability $\alpha$ and ionization potential $I$ \cite{Danielson09,Danielson12,MLbind}, but 
no such accurate universal formula has yet been found. 
Recently we developed an \emph{ab initio} many-body theory (MBT) approach that quantified the role of strong many-body correlations, 
and beyond the interplay of the global properties, highlighted the importance of individual molecular orbital contributions to the positron-molecule potential, e.g., the  enhancement of binding due to $\pi$ bonds \cite{Jaro22,molbind_dft4} that was also deduced from experiment \cite{Danielson09,Ghosh:2022, Danielson:2022} \footnote{It also successfully described positron annihilation on small molecules \cite{Rawlins:2023}}.

Also recently, a model-polarization-potential method \cite{Swann:2018} was used to calculate binding in 
chlorinated hydrocarbons, in a joint theory-experimental study \cite{Swann2021effect}. 
Although
good agreement was found with experiment for many of the molecules considered, for planar molecules the calculations substantially overestimated the measured binding energies, with the suggestion that this was due to the model assuming an isotropic long-range positron-molecule interaction 
\footnote{Specifically, their model constructed the positron-molecule potential as a sum of positron-atom potentials, using a hybrid polarizability of an atom in a certain chemical environment. 
At long range the IPP takes the isotropic asymptotic form $-\alpha/2r^{4}$; while this is true for spherical-top molecules, the true asymptotic form of the polarization potential is generally anisotropic, and is given by $-1/2r^6 \sum x_i x_j \alpha_{ij}$, where the $x_i$ are the Cartesian coordinates and $\alpha_{ij}$ are the Cartesian components of the polarizability tensor \cite{Swann:2018,molbind_dft4}}. 
By contrast, DFT-model calculations 
for planar chloroethylenes \cite{molbind_dft4} accounted for the anisotropy approximately and found better overall agreement with experiment. The method relied on an adjustable gradient parameter $\beta$, whose value the authors of Ref.~\cite{molbind_dft4} were able to chose to replicate the binding energies of dichloroethylenes to within around 10~meV, but this value led to underestimated binding energies for tri- and tetrachloroethylene, at worst by 30 meV.
The anisotropy of the positron-molecule potential, not captured by the global molecular properties, is thus important, and  \emph{ab initio} calculations are demanded for fundamental understanding and description of the body of experimental data.

The purpose of this Letter is twofold. 
First, we apply our many-body theory approach \footnote{We use the fixed-nuclei approximation: compared to the correlations, vibrational effects have been found to have a relatively small effect on the binding energy \cite{Tachikawa14, PhysRevA.73.022705, Romero:2014, Buenker:2007, Buenker:2008,Jaro22}} to study positron binding in the chlorinated hydrocarbons considered in the recent model calculations  \cite{Swann2021effect,molbind_dft4} and experiment \cite{Swann2021effect}, accounting for the positron-molecule correlations and anisotropic potential \emph{ab initio}. We find very good (excellent in cases) agreement with experiment   and DFT-based model calculations, including for the planar molecules.
Secondly, we go beyond the previous chlorinated studies \cite{Swann2021effect,molbind_dft4} and make predictions for fluorinated and brominated molecules, and elucidate the comparative effects of fluorination, chlorination and bromination.
We find that compared to their brominated and chlorinated counterparts, fluorinated molecules generate a successively less attractive positron-molecule potential resulting in very weak or loss of binding.
We identify trends in binding based on global molecular properties ($\alpha$, $\mu$ and $I$) for families including the sequences of cis/(Z)-dihaloethylenes
C$_2$H$_2$Br$_2$ $\to$ 
C$_2$H$_2$BrCl  $\to$ 
C$_2$H$_2$Cl$_2$ $\to$ 
C$_2$H$_2$ClF $\to$ 
C$_2$H$_2$F$_2$ \footnote{Here the descriptor $(Z)$ refers to $(E)$/$(Z)$ isomerism, where $(Z)$ means that the highest priority groups, (i.e. the halogen atoms) are on the same side of the C-C double bond.}, 
and halomethanes. 
However, we find the global properties to be poor universal indicators of binding energies, exemplified by CH$_3$Cl and cis-C$_2$H$_2$F$_2$ which have similar $\alpha$, $\mu$ and $I$ but significantly different positron binding energies (25\,meV vs 3\,meV).
We explain this and the overall results, and provide further fundamental insight by considering the individual MO contributions to the positron-molecule correlation potential, showing that e.g., the decrease (or loss of) binding for bromination$\to$chlorination$\to$fluorination is due to successively higher molecular orbital ionization energies and smaller density of states  close to the HOMO.

\begin{figure}[t!]
		\includegraphics[width=0.44\textwidth]{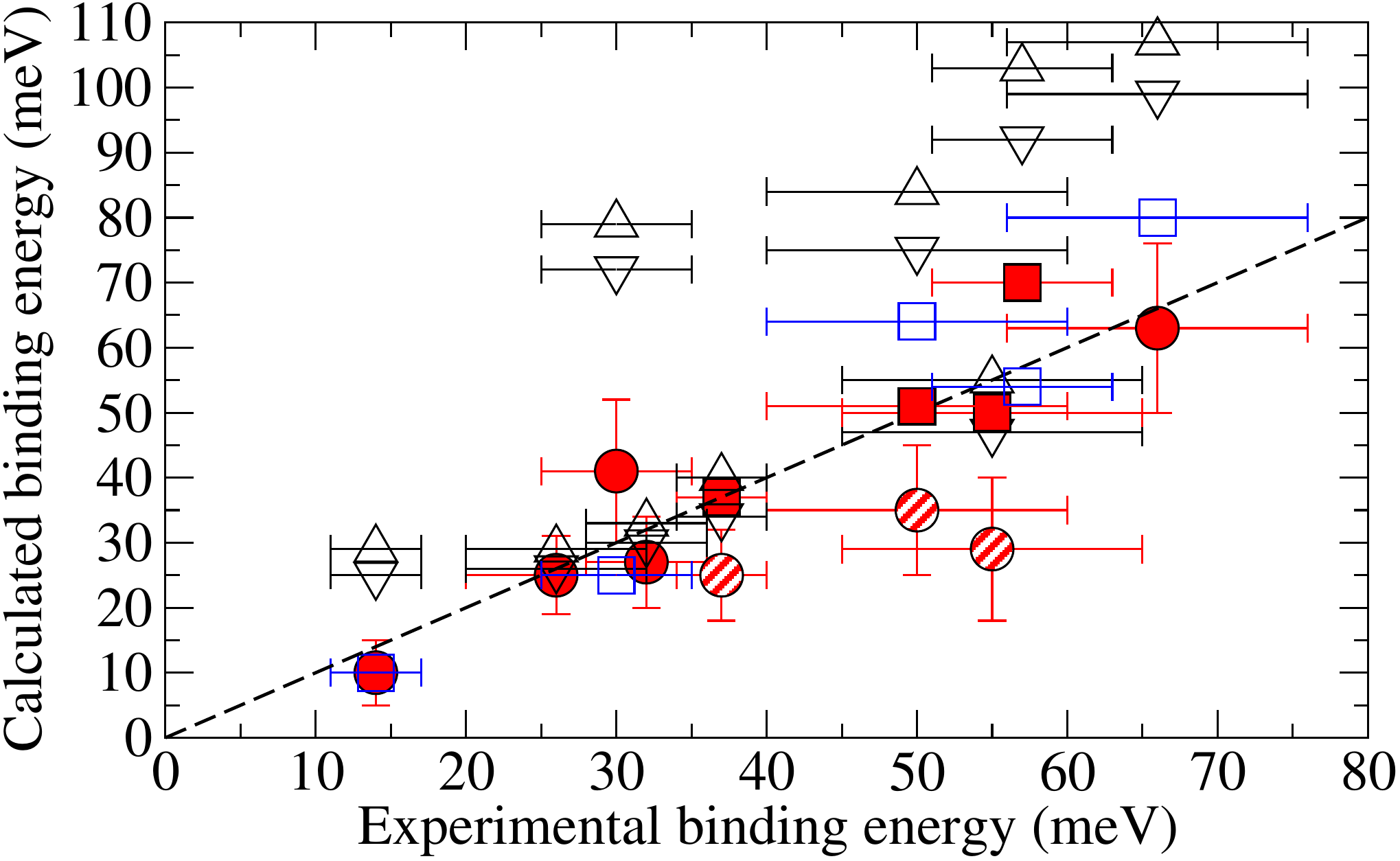}
	\includegraphics[width=0.45\textwidth]{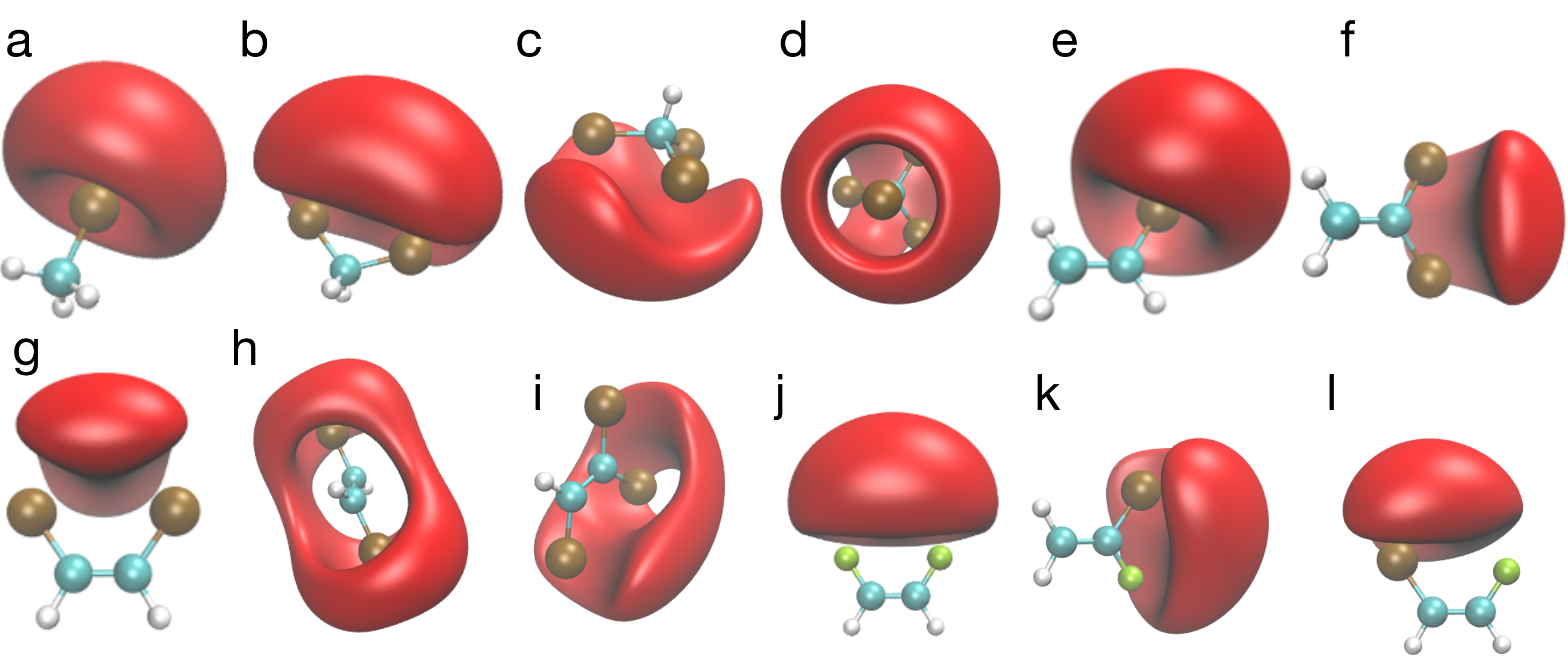}
	\vspace*{-2mm}
	\caption{Top: calculated positron binding energies compared with experiment for chlorinated molecules: 
present MBT (red circles, and striped circles for molecules difficult to converge \emph{ab initio}); 
MBT-based model calculations using $\Sigma=g\Sigma^{(2)} + \Sigma^{\Lambda}$ with $g=1.5$ (red squares);  
(isotropic) polarization potential model calculation of \cite{Swann2021effect} (black triangles; two for each molecule reflecting two choices of cut-off parameter); 
DFT-model calculation of \cite{molbind_dft4} (blue squares). 
Vertical error bars are plus-minus the maximum difference of our calculations using screened Coulomb interactions and $GW$@RPA MO energies vs bare Coulomb interactions and HF MO energies \cite{Jaro22}.
Bottom: positron (Dyson) wave function at 80\% maximum for chlorinated and fluorinated molecules with $\varepsilon_b\geq1$ meV; a)~Chloromethane; b)~Dichloromethane; c)~Trichloromethane; d)~Tetrachloromethane (at 93\%); e)~Vinyl chloride; f)~Vinylidene chloride; g)~\textit{cis}-1,2-dichloroethylene; h)~\textit{trans}-1,2-dichloroethylene (at 90\%); i)~Trichloroethylene; j)~\textit{cis}-1,2-difluoroethylene; k)~1-chloro-1-fluoroethylene; l)~(\textit{Z})-chlorofluoroethylene.}
	\label{fig calc vs exp and orbitals}
\end{figure}

\textit{Theoretical approach.---}A detailed description of our 
MBT approach is given in \cite{Jaro22}. 
Briefly, we solve the Dyson equation \cite{mbtexposed,fetterwalecka} $(\hat{H}_0+\hat{\Sigma}_\varepsilon)\psi_\varepsilon(\mathbf{r})=\varepsilon\psi_{\varepsilon}(\mathbf{r})$
 self-consistently for the positron wave function $\psi_\varepsilon(\mathbf{r})$ with energy $\varepsilon$. Here $\hat{H}_0$ is the zeroth-order Hamiltonian of the positron in the static (Hartree-Fock) field of the molecule and $\hat{\Sigma}_\varepsilon$ is the positron self energy (an energy-dependent, non-local correlation potential) \cite{PhysRevLett.3.96}. 
 We calculate it using a diagrammatic expansion in electron-electron and electron-positron interactions, see Fig.~1 
 of \cite{Jaro22}, involving three main diagram classes: the \textit{GW} diagram, which describes polarization, screening of the electron-positron Coulomb interaction, and electron-hole interactions; the virtual-positronium (vPs) formation ladder series, which describes the temporary tunnelling of an electron to the positron, denoted $\Sigma^{\Gamma}$; and the positron-hole repulsion ladder series, denoted $\Sigma^{\Lambda}$
The significant enhancement and enabling of binding due to these correlations were delineated in \cite{Jaro22}. Here we quote results only for our most sophisticated self-energy $\Sigma^{GW+\Gamma+\Lambda}$  \footnote{We use screened Coulomb interactions in the ladder diagrams 
and MO energies calculated in the random phase approximation. The $\varepsilon_b$ in this approximation are found to be within 4~meV of those calculated using bare Coulomb interactions and HF MO energies for chlorinated and brominated molecules (maximum relative error of $7.4\%$), and 0.1~meV for fluorinated molecules.}. 
We expand the electron and positron wave functions in Gaussian basis sets, using aug-cc-pVXZ bases (X=T,Q) \cite{Dunning} on atomic centres as well as additional hydrogen aug-cc-pVXZ bases on ``ghost'' centres 1${\rm{\AA}}$\phantom{} from the molecule to resolve regions of maximum positron density. For all of the molecules considered, we placed 5 ghosts around each halogen atom in the molecule in the shape of a square-pyramidal cap, with each ghost 1\AA\phantom{} from the halogen (see Supplemental Material ``SM"). We also use diffuse even-tempered positron bases of the form $10s9p8d7f6g$, with exponents $\zeta_0\times\beta^{k-1}$  ($\zeta_0=0.00001$-- $0.006$ and $\beta=2$-- $3$), ensuring the positron is described well at large distances $r\sim1/\kappa$, where $\kappa=\sqrt{2\varepsilon_b}$.
For molecules with $>2$ chlorines the positron wave function is delocalized (Fig.~\ref{fig calc vs exp and orbitals}), and we found that accurate description of the vPs contribution 
requires a prohibitively large basis set  \footnote{The vPs contribution to the self energy requires diagonalization of dense matrices of size $(N_+N_-)^2$, where $N_+$ ($N_-$) is the number of virtual positron (electron) states used. For example, our C$_2$HCl$_3$ calculations have $N_+N_-=280800$, corresponding to $630$ GB of memory. With other memory costs considered, our approach regularly requires more than 1 TB of memory.}
(for our current computing resources), and our \emph{ab initio} calculations are not converged, though are lower bounds. 
Thus we also performed MBT-based model calculations approximating $\Sigma\approx g\Sigma^{(2)}+\Sigma^{(\Lambda)}$, using the second-order self-energy scaled to approximate the virtual-Ps contribution as introduced and justified in \cite{Jaro22}: \emph{ab initio} calculations give $g$ in the range 1.4 to 1.5 for the HOMOs [see \cite{Jaro22} and also Fig.~2 (d).]
This approach still calculates the anisotropic polarization potential \emph{ab initio}, but is much less computationally expensive.

\textit{Chlorinated molecules: comparison with experiment and model calculations.---}Our calculated positron binding energies $\varepsilon_b$ for the chlorinated hydrocarbons considered in the recent isotropic-polarization-potential (IPP) \cite{Swann2021effect} and DFT model calculations  \cite{molbind_dft4} and experiment \cite{Swann2021effect}, and our predictions for their fluorinated and (select) brominated counterparts are presented in Table~\ref{tab:bind}. 
Figure~\ref{fig calc vs exp and orbitals} summarizes this for the chlorinated molecules, and also presents the calculated bound-state positron Dyson orbitals for chlorinated and select chloro-fluorinated molecules, showing that the positron localizes around the halogens. 
Overall, very good agreement is found between the \emph{ab initio} MBT calculations and experiment.
For CH$_3$Cl, our calculated $\varepsilon_b=25$ meV
is in excellent agreement with both experiment and the IPP model calculations.
We find excellent agreement with experiment for CH$_2$Cl$_2$,  and for \textit{cis}-C$_2$H$_2$Cl$_2$ (for which both the IPP and DFT models substantially overestimate) and \textit{trans}-C$_2$H$_2$Cl$_2$, and reasonable agreement for vinylidene chloride C$_2$H$_2$Cl$_2$. 
Overall, our \textit{ab initio} results are in good agreement with the DFT-based calculations \cite{molbind_dft4} (including vinyl chloride, for which there is no measurement).
The results of the MBT-based model calculation, which importantly augment our unconverged \emph{ab initio} results for the molecules with $>2$ chlorines, are presented in the final column of Table \ref{tab:bind}. 
The model calculations with $g\sim1.5$ generally give excellent agreement with experiment (with the exception of ethylene). 

\begin{table*}
	\caption{\label{tab:bind}
		Calculated MBT positron binding energies (meV) for halogenated hydrocarbons compared with experiment and model-potential calculations. For calculations denoted `$<0$' binding was not observed. Where $\varepsilon_b <1$ meV, we quote values to 1 decimal place. Molecules marked `*' are those for which we believe our \textit{ab initio} calculations to be unconverged and we recommend the model-MBT result (final column, see text). Also shown are calculated HF dipole moments, isotropic dipole polarizabilities (calculated at the $GW$@BSE level) and ionization energies (calculated at the $GW$@RPA level and used in the energy denominators of the self-energy analytic expressions \cite{Jaro22}). 
		}
	\begin{ruledtabular}
		\begin{tabular}{l@{\hskip4pt}l@{\hskip6pt}c@{\hskip4pt} c@{\hskip4pt}c@{\hskip8pt} c@{\hskip4 pt}c@{\hskip4pt} c@{\hskip4pt} c@{\hskip12pt}c@{\hskip12pt}  }
			&& &  & & Present \emph{ab initio} MBT & &\multicolumn{3}{c}{Model-potential calculations}  \\			
			\cline{6-6} \cline{8-10} \\[-1.5ex] 
			Molecule& Formula &$\mu$\,(D) &$\alpha$\,(a.u.) &$I$\,(eV) & $\Sigma^{GW+\Gamma+\Lambda}$	& Exp. \cite{Swann2021effect} & IPP\footnote{Model-polarization-potential calculations of Swann and Gribakin, assuming isotropic asymptotic interaction \cite{Swann2021effect}.} 	& DFT \footnote{DFT is the density-functional theory using the Perdew-Burke-Ernzerhof exchange functional  result from Suzuki \emph{et al.} \cite{molbind_dft4}.}	& \multicolumn{1}{c}{Present MBT-based model\footnote{Using a scaled self-energy $\Sigma=g\Sigma^{(2)}+\Sigma^{(\Lambda)}$ with $g$ ranging from 1.4 to 1.5 to account for vPs formation \cite{Jaro22}.}}   \\

			\hline\\[-1ex]
			Methane 					&  CH$_4$ 			& 0 & 13.83 &  	14.18 & $<0$			&   --
				&  $<0$		& -- &   $<0$ \\
			Chloromethane 					& CH$_3$Cl 			&2.15 &27.80 & 11.78 &\textbf{25}			& $26\pm6$			& 29, 26		&-- & 8--23  \\
			Dichloromethane 				& CH$_2$Cl$_2$ 			& 1.83  &40.61& 	11.93 &\textbf{27}	& $32\pm$4		& 34, 30		&-- & 15--31   \\
			Trichloromethane$^{\ast}$			& CHCl$_3$ 		& 1.19 & 53.34  & 		11.95	&	{25}$^{\ast}$	& $37\pm$3		& 40, 34		&--  & 16--{\bf 37}  \\
			Tetrachloromethane$^{\ast}$				& CCl$_4$ 			&0&  64.14  & 		12.02	&  {29}$^{\ast}$	& $55\pm$10		& 55, 47		&--  & 22--{\bf 50}  \\
			Ethylene 					& C$_2$H$_4$ 			&0 &24.40 & 	10.75 &1			&  $20\pm10$			& 5		&-- &  $<0$   \\
			Vinyl chloride			&  	C$_2$H$_3$Cl		&  1.68   &   38.67    & 	10.57& \textbf{27}	& 	--	& 54, 50		&27  & 8--28   \\		
			Vinylidene chloride				&  C$_2$H$_2$Cl$_2$	&  1.62  &  51.04  & 		10.50	&\textbf{41} &	$30\pm$5	&	79, 72	&25  & 13--30    \\ 
			\textit{cis}-1,2-dichloroethylene			& C$_2$H$_2$Cl$_2$	&  2.13  &  51.18  & 	10.34	&\textbf{63	}	&	$66\pm$10	&	107, 99	&80  & 43--75  \\ 			
			\textit{trans}-1,2-dichloroethylene 				& C$_2$H$_2$Cl$_2$			&0&  52.79  & 		10.27	 &\textbf{10}	&	$14\pm$3	&	29, 25	&10   & 2--12   \\
			Trichloroethylene$^{\ast}$			& C$_2$HCl$_3$ 			&  1.01  &  64.84  & 	10.16	&	{35}$^{\ast}$		&	$50\pm$10	&	84, 75	&64  &  23--{\bf 51}   \\
			Tetrachloroethylene			& C$_2$Cl$_4$ 			&  0  &  87.02 & 9.46	&	--		&	$57\pm$6	&	103, 92	&54  &  34--70  \\[1ex]
			
			1-chloro-1-fluoroethylene			& C$_2$H$_2$ClF 			&  1.49  &  38.46  & 	10.71&	\textbf{5}	&	--	&	--	&--  &  2--10   \\
			(\textit{Z})-chlorofluoroethylene			& C$_2$H$_2$ClF 			&  2.37  &  38.52  & 	10.53 &	 \textbf{32}	&	--	&	--	&--  &  22--39   \\[1ex]

			Fluoromethane 					& CH$_3$F			&  1.94 &   15.56  & 	13.99	& 	\textbf{0.3}	& 0.3\footnote{Molecule is VFR active, but $\varepsilon_b$ is too small to measure \cite{Young08_small}. 0.3 meV was derived from the Z$_{\text{eff}}$ fit of the VFR-based annihilation spectrum \cite{Gleb2006}.  }			& 	--	&--  & 0.2--0.6  \\
			Difluoromethane 				& CH$_2$F$_2$ 			&  2.09  &  16.15  & 		13.70	& \textbf{0.2}	& 	--	& --		&--  & 0.1--0.3  \\
			Trifluoromethane			& CHF$_3$ 		&  1.75  &  16.66 & 		15.17	&	$<0$	&--		& 	--	&--  & --   \\
			Tetrafluoromethane 				& CF$_4$ 			& 0 &  17.13  & 		16.26	& 	$<0$	& \phantom{0}-- \footnote{CF$_{4}$ is not VFR active \cite{Young08_small}.}		& 	--	&--  & --  \\
			Vinyl fluoride			& C$_2$H$_3$F 			& 1.47 &  26.11  & 	10.92  &	\textbf{0.3}& --	& --	&--   & 0--0.6    \\
			Vinylidene fluoride				& C$_2$H$_2$F$_2$ 			&  1.30  &  26.22  & 		10.88	&		$<0$	&	--	&	--	&--   & --    \\
			\textit{cis}-1,2-difluoroethylene 			& C$_2$H$_2$F$_2$		&  2.49 &  26.50  & 		10.73 &		\textbf{3}	&	--	&	--	&--   & 1--7 \\
			\textit{trans}-1,2-difluoroethylene 				& C$_2$H$_2$F$_2$ 			& 0 &  26.25  & 		10.68	&		$<0$	&	--	&	--	&--  & --   \\
			Trifluoroethylene			& C$_2$HF$_3$ 			&  1.37  &  26.43  & 	10.75	& $<0$ &		--	&	--	&	--	&	--	 \\[1ex]
			
			Bromomethane			& CH$_3$Br 			&  2.18  &  34.75  & 	10.93&	\textbf{56}	&	40\footnote{From Ref.~\cite{Young08_small}, where the uncertainty in the $Z_{\text{eff}}$ peak positions from which $\varepsilon_b$ was measured was reported to be between 10 and 15 meV.}	&	--	&--  &  23--41   \\
			\textit{cis}-1,2-dibromoethylene			& C$_2$H$_2$Br$_2$			&  1.97  &  64.67  & 	10.09 &	\textbf{109} 		&	--	&	--	&--  &  58--108   \\
			(\textit{Z})-bromochloroethylene			& C$_2$H$_2$BrCl			&  2.04  &  57.72  & 	10.20 & \textbf{80 }	&	--	&	--	&--  &  42--87  
		\end{tabular}
	\end{ruledtabular}
\end{table*}

\begin{figure*}[ht!]
	\includegraphics[width=\textwidth]{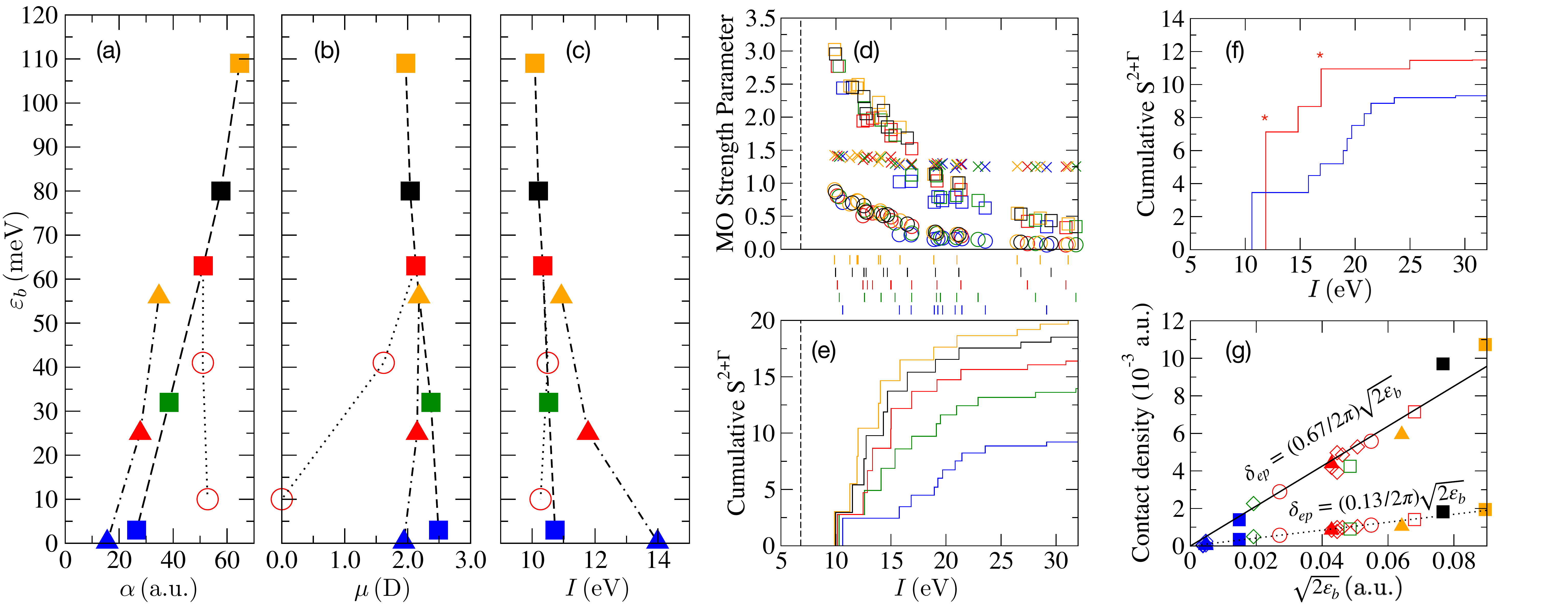}
	\vspace*{-5mm}
	\caption{Dependence of positron binding energies on global molecular properties and individual MOs. (a)-(c): calculated $\varepsilon_b$ vs. calculated polarizabilities, dipole moments and ionization energy for the brominated (orange), bromochlorinated (black), chlorinated (red), chlorofluorinated (green) and fluorinated (blue) molecules; symbols denote molecular families: squares are \textit{cis}-dihaloethylenes C$_2$H$_2$XY, triangles are halomethanes CH$_3$X (X,Y = Br, Cl, F) and circles are isomers of C$_2$H$_2$Cl$_2$. Dashed lines are guides; (d) the positron-molecule correlation strength parameters $\mathcal{S}_n^{\Gamma}$ (circles) and $\mathcal{S}_n^{2+\Gamma}$ (squares), and the ratio $g_n\equiv\mathcal{S}_n^{2+\Gamma}/\mathcal{S}_n^2$ (crosses) for each MO $n$ against the MO HF ionization energies (vertical lines between panels) for the \textit{cis}-dihaloethylenes sequence [colours as in (a)-(c)]. (e) the corresponding cumulative $\mathcal{S}^{2+\Gamma}$ obtained by summing from the HOMO to the core orbitals. (f)  the cumulative strength $\mathcal{S}^{2+\Gamma}$ for CH$_3$Cl (red; asterisks denote double degeneracy) and \textit{cis}-C$_2$H$_2$F$_2$ (blue). (g) the calculated unenhanced ($\gamma_i=1$) and enhanced contact densities for molecules with a $\Sigma^{GW+\Gamma+\Lambda}$ bound state. Colours and symbols as in (a)-(c), diamonds are remaining molecules from Table~\ref{tab:bind}.}
\label{fig strength param}
\end{figure*}

\textit{Fluorinated molecules: predictions.---}Compared to the chlorinated molecules, in the fluorinated counterparts we find (see Table \ref{tab:bind}) that positron binding is either lost or greatly reduced (as explained in the next section). We predict bound states for fluoromethane, difluoroethylene, vinyl fluoride (a few tenths of a meV each) and \textit{cis}-1,2-difluoroethylene ($\varepsilon_b\sim3$ meV). 
Although fluoromethane is known to be VFR active, $\varepsilon_b$ was found to be too small to measure \cite{Young08_small}. 
However, our prediction of a weak bound-state for fluoromethane of $\sim0.3$ meV is in agreement with that derived from the Z$_{\text{eff}}$ fit of the annihilation spectrum of CH$_3$F, which until now had not been corroborated with any theoretical calculations \cite{Gleb2006}. This contradicts a recent machine-learning-based prediction that fluoromethane does not bind a positron \cite{MLbind}. Our prediction of a bound state for CH$_2$F$_2$ with $\varepsilon_b=0.2$ meV concurs with the $0.4$ meV prediction by an earlier empirical model \cite{Danielson09}. Our lack of binding in CF$_4$ is consistent with experiment; this molecule is known to not be VFR active \cite{Young08_small}. We also considered 1-chloro-1-fluoroethylene and (\textit{Z})-chlorofluoroethylene, and report binding energies of 5 meV and 32 meV. These values lie between the fully chlorinated and fluorinated binding energies (see below). 

\emph{Comparative effect of fluorination, chlorination and bromination; the role of MO energies and density of states.---}{Figure~\ref{fig strength param}} (a)-(c) show the calculated $\varepsilon_b$ as a function of the global molecular properties $\alpha$, $\mu$, and $I$  for the dihaloethylenes (\textit{cis}/\textit{Z}-C$_2$H$_2$XY and the isomers of C$_2$H$_2$Cl$_2$) and halomethanes CH$_3$X, where X,Y$=$ F, Cl, or Br. 
These present three distinct cases. 
Across the \textit{cis}-dihaloethylenes $I$ and $\mu$ vary weakly, and the increase in $\varepsilon_b$ going from X,Y$=$ F$_2$ $\to$ ClF $\to$ ... $\to$ Br$_2$ follows an increase in $\alpha$: in a given family a more polarizable target is more attractive to the positron.
Across the halomethanes, $\mu$ is almost constant, and the increase in $\varepsilon_b$ going from F to Br follows both an increasing $\alpha$ and decreasing $I$ (the less tightly bound electrons are more susceptible to perturbation from the positron). 
For the isomers of C$_2$H$_2$Cl$_2$, $\alpha$ and $I$ are vary weakly, and the decrease in $\varepsilon_b$ from \textit{cis}-C$_2$H$_2$Cl$_2$ to vinylidene chloride to the non-polar \textit{trans}-C$_2$H$_2$Cl$_2$ is due to successively decreasing $\mu$. 
These three distinct cases highlight that the global molecular properties can explain trends in $\varepsilon_b$ for families of molecules, but they are not reliable universal predictors of binding energies, as exemplified by considering CH$_3$Cl and \textit{cis}-C$_2$H$_2$F$_2$. These have very similar $\alpha$, but whilst \textit{cis}-C$_2$H$_2$F$_2$ has a larger $\mu$ and lower $I$, it has a lower binding energy (3 meV vs.~25 meV). 
To explain this, and the reduction or lack of binding in fluorinated molecules in general, we consider
the individual molecular orbital contributions to the correlation potential. 
We do so via the strength parameter $\mathcal{S}=-\sum_{\nu}\bra{\nu}\hat{\Sigma}_\varepsilon\ket{\nu}/\varepsilon_\nu$ \cite{Dzuba:1994,Jaro22}, where $\nu$ is an excited positron Hartree-Fock (HF) orbital of energy $\varepsilon_\nu$, with the self energy taken as $\Sigma\approx \Sigma^{(2+\Gamma)}$, i.e., the sum of the bare polarization $\Sigma^{(2)}$ and the virtual-Ps  $\Sigma^{(\Gamma)}$ diagrams. 
Figure \ref{fig strength param} (d) shows $\mathcal{S}^{(\Gamma)}$,  $\mathcal{S}^{(2+\Gamma)}$ and the ratio $g=\mathcal{S}^{(2+\Gamma)}/\mathcal{S}^{(2)}$
for individual MOs as a function of the MO energy for the sequence of \textit{cis}-dihaloethylenes: the strength parameters mainly decrease with increasing MO ionization energy because more tightly bound orbitals are more difficult for the positron to perturb \cite{Jaro22}. 
 Additionally, Fig.~2 (e) shows the cumulative $\mathcal{S}^{(2+\Gamma)}$ obtained by summing from the HOMO to the core orbitals.
Moving from C$_2$H$_2$Br$_2$ through to C$_2$H$_2$F$_2$ sees both the total $\mathcal{S}^{(2+\Gamma)}$ and the density of states near the ionization energy decrease: e.g., in C$_2$H$_2$F$_2$ there is a $\sim5$ eV gap between the HOMO and the HOMO$-1$, while this gap is approximately half as wide for C$_2$H$_2$Cl$_2$ and C$_2$H$_2$ClF and half as wide again for C$_2$H$_2$Br$_2$. 
Further, the contributions to the cumulative $\mathcal{S}^{(2+\Gamma)}$ below the HOMO for C$_2$H$_2$F$_2$ are smaller than those for the other three molecules as the MOs have larger $I$. In general the transition from Br to Cl to F either shifts all the energy states to more negative energies, or at least drives the sub-HOMO energies further from the HOMO energy, inhibiting the molecule's ability to bind the positron (SM Fig.~S1 shows MO energies of all molecules considered). 
We now consider CH$_3$Cl and \textit{cis}-C$_2$H$_2$F$_2$ [red triangle and blue square in Fig.~\ref{fig strength param} (a)-(c)].
Figure~\ref{fig strength param} (f) shows their cumulative $\mathcal{S}^{(2+\Gamma)}$ strength parameter.
We see that although CH$_3$Cl has a larger $I$, its HOMO is doubly degenerate, and contributes relatively more to the strength than the singly-degenerate HOMO of CH$_2$F$_2$ (a second doubly degenerate state of $\pi$ character also contributes strongly at $\sim 17$\,eV for CH$_3$Cl). Thus, in spite of CH$_3$Cl having a smaller dipole moment (which governs the strength of the static potential \cite{Gribakin2015}), its larger correlation potential (which contributes to binding non-linearly; see Extended Data Fig.~{3} of \cite{Jaro22}) ultimately results in stronger binding.

\textit{Annihilation and contact densities.---}
The positron bound state annihilation rate 
$\Gamma {\rm[ns^{-1}]} = 50.47\, \delta_{ep}{\rm [a.u.]}$, where 
$\delta_{ep}=\sum_{i=1}^{N_e/2}\gamma_i\int\lvert\varphi_i(\mathbf{r})\rvert^2\lvert\psi_{\varepsilon}(\mathbf{r})\rvert^2d^3\mathbf{r}$ is the electron-positron contact density. Here $\varphi_i$ is the $i$-th electron MO, $\psi$ is the positron bound-state wavefunction (see e.g., Fig.~1) renormalized to $a=\left(1-\partial\varepsilon/\partial E|_{\varepsilon_b} \right)^{-1}<1$ \cite{Jaro22} and $\gamma_i\geq 1$ are vertex enhancement factors that account for 
short-range electron-positron attractions \cite{DGG:2015:core,DGG:2017:ef}. 
We found that they  
followed $\delta_{ep}=(F/2\pi)\sqrt{2\varepsilon_b}$ \cite{Gribakin2001} 
with $F=0.67$,
remarkably close to $F\approx0.66$ for atoms \cite{Gribakin2001}, see Fig.~2 (g).

{\textit{Summary.---}Many-body theory calculations of positron binding to chlorinated hydrocarbons were found to be in good to excellent agreement with experiment and recent model-potential-based DFT calculations. 
Additionally, new predictions elucidated the comparative effects of fluorination, chlorination and bromination: 
trends within molecular families 
based on the global molecular properties $\mu$, $\alpha$ and $I$ were identified, as was
the importance of describing the positron-molecule potential anisotropy, and accounting for the energies and density of electron states (at least near the HOMO).
We suggest that any accurate universal formula for positron binding energies should thus include these molecular properties. As well as providing fundamental insight, our results provide benchmarks and can inform other computational approaches to the positron-molecule and many-electron problems. 
}

\begin{acknowledgments}
	\emph{Acknowledgements}.---We thank Sarah Gregg, Gleb Gribakin and Andrew Swann for useful discussions. This work was funded by the European Research Council grant 804383 `ANTI-ATOM' 
	and a DfE Northern Ireland postgraduate research studentship (JPC), and used the NI HPC Service and the ARCHER2 UK National Supercomputing Service. 
\end{acknowledgments}

%


\begin{thebibliography}{42}%
\makeatletter
\providecommand \@ifxundefined [1]{%
 \@ifx{#1\undefined}
}%
\providecommand \@ifnum [1]{%
 \ifnum #1\expandafter \@firstoftwo
 \else \expandafter \@secondoftwo
 \fi
}%
\providecommand \@ifx [1]{%
 \ifx #1\expandafter \@firstoftwo
 \else \expandafter \@secondoftwo
 \fi
}%
\providecommand \natexlab [1]{#1}%
\providecommand \enquote  [1]{``#1''}%
\providecommand \bibnamefont  [1]{#1}%
\providecommand \bibfnamefont [1]{#1}%
\providecommand \citenamefont [1]{#1}%
\providecommand \href@noop [0]{\@secondoftwo}%
\providecommand \href [0]{\begingroup \@sanitize@url \@href}%
\providecommand \@href[1]{\@@startlink{#1}\@@href}%
\providecommand \@@href[1]{\endgroup#1\@@endlink}%
\providecommand \@sanitize@url [0]{\catcode `\\12\catcode `\$12\catcode
  `\&12\catcode `\#12\catcode `\^12\catcode `\_12\catcode `\%12\relax}%
\providecommand \@@startlink[1]{}%
\providecommand \@@endlink[0]{}%
\providecommand \url  [0]{\begingroup\@sanitize@url \@url }%
\providecommand \@url [1]{\endgroup\@href {#1}{\urlprefix }}%
\providecommand \urlprefix  [0]{URL }%
\providecommand \Eprint [0]{\href }%
\providecommand \doibase [0]{https://doi.org/}%
\providecommand \selectlanguage [0]{\@gobble}%
\providecommand \bibinfo  [0]{\@secondoftwo}%
\providecommand \bibfield  [0]{\@secondoftwo}%
\providecommand \translation [1]{[#1]}%
\providecommand \BibitemOpen [0]{}%
\providecommand \bibitemStop [0]{}%
\providecommand \bibitemNoStop [0]{.\EOS\space}%
\providecommand \EOS [0]{\spacefactor3000\relax}%
\providecommand \BibitemShut  [1]{\csname bibitem#1\endcsname}%
\let\auto@bib@innerbib\@empty
\bibitem [{\citenamefont {Gilbert}\ \emph {et~al.}(2002)\citenamefont
  {Gilbert}, \citenamefont {Barnes}, \citenamefont {Sullivan},\ and\
  \citenamefont {Surko}}]{Gilbert02}%
  \BibitemOpen
  \bibfield  {author} {\bibinfo {author} {\bibfnamefont {S.~J.}\ \bibnamefont
  {Gilbert}}, \bibinfo {author} {\bibfnamefont {L.~D.}\ \bibnamefont {Barnes}},
  \bibinfo {author} {\bibfnamefont {J.~P.}\ \bibnamefont {Sullivan}},\ and\
  \bibinfo {author} {\bibfnamefont {C.~M.}\ \bibnamefont {Surko}},\ }\bibfield
  {title} {\bibinfo {title} {Vibrational-{R}esonance {E}nhancement of
  {P}ositron {A}nnihilation in {M}olecules},\ }\href
  {https://doi.org/10.1103/PhysRevLett.88.043201} {\bibfield  {journal}
  {\bibinfo  {journal} {Phys. Rev. Lett.}\ }\textbf {\bibinfo {volume} {88}},\
  \bibinfo {pages} {043201} (\bibinfo {year} {2002})}\BibitemShut {NoStop}%
\bibitem [{\citenamefont {Gribakin}\ \emph {et~al.}(2010)\citenamefont
  {Gribakin}, \citenamefont {Young},\ and\ \citenamefont
  {Surko}}]{RevModPhys.82.2557}%
  \BibitemOpen
  \bibfield  {author} {\bibinfo {author} {\bibfnamefont {G.~F.}\ \bibnamefont
  {Gribakin}}, \bibinfo {author} {\bibfnamefont {J.~A.}\ \bibnamefont
  {Young}},\ and\ \bibinfo {author} {\bibfnamefont {C.~M.}\ \bibnamefont
  {Surko}},\ }\bibfield  {title} {\bibinfo {title} {Positron-molecule
  interactions: Resonant attachment, annihilation, and bound states},\ }\href
  {https://doi.org/10.1103/RevModPhys.82.2557} {\bibfield  {journal} {\bibinfo
  {journal} {Rev. Mod. Phys.}\ }\textbf {\bibinfo {volume} {82}},\ \bibinfo
  {pages} {2557} (\bibinfo {year} {2010})}\BibitemShut {NoStop}%
\bibitem [{\citenamefont {Danielson}\ \emph {et~al.}(2009)\citenamefont
  {Danielson}, \citenamefont {Young},\ and\ \citenamefont
  {Surko}}]{Danielson09}%
  \BibitemOpen
  \bibfield  {author} {\bibinfo {author} {\bibfnamefont {J.~R.}\ \bibnamefont
  {Danielson}}, \bibinfo {author} {\bibfnamefont {J.~A.}\ \bibnamefont
  {Young}},\ and\ \bibinfo {author} {\bibfnamefont {C.~M.}\ \bibnamefont
  {Surko}},\ }\bibfield  {title} {\bibinfo {title} {Dependence of
  positron-molecule binding energies on molecular properties},\ }\href
  {https://doi.org/10.1088/0953-4075/42/23/235203} {\bibfield  {journal}
  {\bibinfo  {journal} {J. Phys. B}\ }\textbf {\bibinfo {volume} {42}},\
  \bibinfo {pages} {235203} (\bibinfo {year} {2009})}\BibitemShut {NoStop}%
\bibitem [{\citenamefont {Danielson}\ \emph {et~al.}(2010)\citenamefont
  {Danielson}, \citenamefont {Gosselin},\ and\ \citenamefont
  {Surko}}]{Danielson10}%
  \BibitemOpen
  \bibfield  {author} {\bibinfo {author} {\bibfnamefont {J.~R.}\ \bibnamefont
  {Danielson}}, \bibinfo {author} {\bibfnamefont {J.~J.}\ \bibnamefont
  {Gosselin}},\ and\ \bibinfo {author} {\bibfnamefont {C.~M.}\ \bibnamefont
  {Surko}},\ }\bibfield  {title} {\bibinfo {title} {Dipole {E}nhancement of
  {P}ositron {B}inding to {M}olecules},\ }\href
  {https://doi.org/10.1103/PhysRevLett.104.233201} {\bibfield  {journal}
  {\bibinfo  {journal} {Phys. Rev. Lett.}\ }\textbf {\bibinfo {volume} {104}},\
  \bibinfo {pages} {233201} (\bibinfo {year} {2010})}\BibitemShut {NoStop}%
\bibitem [{\citenamefont {Danielson}\ \emph
  {et~al.}(2012{\natexlab{a}})\citenamefont {Danielson}, \citenamefont {Jones},
  \citenamefont {Gosselin}, \citenamefont {Natisin},\ and\ \citenamefont
  {Surko}}]{Danielson12}%
  \BibitemOpen
  \bibfield  {author} {\bibinfo {author} {\bibfnamefont {J.~R.}\ \bibnamefont
  {Danielson}}, \bibinfo {author} {\bibfnamefont {A.~C.~L.}\ \bibnamefont
  {Jones}}, \bibinfo {author} {\bibfnamefont {J.~J.}\ \bibnamefont {Gosselin}},
  \bibinfo {author} {\bibfnamefont {M.~R.}\ \bibnamefont {Natisin}},\ and\
  \bibinfo {author} {\bibfnamefont {C.~M.}\ \bibnamefont {Surko}},\ }\bibfield
  {title} {\bibinfo {title} {Interplay between permanent dipole moments and
  polarizability in positron-molecule binding},\ }\href
  {https://doi.org/10.1103/PhysRevA.85.022709} {\bibfield  {journal} {\bibinfo
  {journal} {Phys. Rev. A}\ }\textbf {\bibinfo {volume} {85}},\ \bibinfo
  {pages} {022709} (\bibinfo {year} {2012}{\natexlab{a}})}\BibitemShut
  {NoStop}%
\bibitem [{\citenamefont {Danielson}\ \emph
  {et~al.}(2012{\natexlab{b}})\citenamefont {Danielson}, \citenamefont {Jones},
  \citenamefont {Natisin},\ and\ \citenamefont {Surko}}]{Danielson12a}%
  \BibitemOpen
  \bibfield  {author} {\bibinfo {author} {\bibfnamefont {J.~R.}\ \bibnamefont
  {Danielson}}, \bibinfo {author} {\bibfnamefont {A.~C.~L.}\ \bibnamefont
  {Jones}}, \bibinfo {author} {\bibfnamefont {M.~R.}\ \bibnamefont {Natisin}},\
  and\ \bibinfo {author} {\bibfnamefont {C.~M.}\ \bibnamefont {Surko}},\
  }\bibfield  {title} {\bibinfo {title} {Comparisons of {P}ositron and
  {E}lectron {B}inding to {M}olecules},\ }\href
  {https://doi.org/10.1103/PhysRevLett.109.113201} {\bibfield  {journal}
  {\bibinfo  {journal} {Phys. Rev. Lett.}\ }\textbf {\bibinfo {volume} {109}},\
  \bibinfo {pages} {113201} (\bibinfo {year} {2012}{\natexlab{b}})}\BibitemShut
  {NoStop}%
\bibitem [{\citenamefont {Swann}\ \emph {et~al.}(2021)\citenamefont {Swann},
  \citenamefont {Gribakin}, \citenamefont {Danielson}, \citenamefont {Ghosh},
  \citenamefont {Natisin},\ and\ \citenamefont {Surko}}]{Swann2021effect}%
  \BibitemOpen
  \bibfield  {author} {\bibinfo {author} {\bibfnamefont {A.~R.}\ \bibnamefont
  {Swann}}, \bibinfo {author} {\bibfnamefont {G.~F.}\ \bibnamefont {Gribakin}},
  \bibinfo {author} {\bibfnamefont {J.~R.}\ \bibnamefont {Danielson}}, \bibinfo
  {author} {\bibfnamefont {S.}~\bibnamefont {Ghosh}}, \bibinfo {author}
  {\bibfnamefont {M.~R.}\ \bibnamefont {Natisin}},\ and\ \bibinfo {author}
  {\bibfnamefont {C.~M.}\ \bibnamefont {Surko}},\ }\bibfield  {title} {\bibinfo
  {title} {Effect of chlorination on positron binding to hydrocarbons:
  experiment and theory},\ }\href {https://doi.org/10.1103/PhysRevA.104.012813}
  {\bibfield  {journal} {\bibinfo  {journal} {Phys. Rev. A}\ }\textbf {\bibinfo
  {volume} {104}},\ \bibinfo {pages} {012813} (\bibinfo {year}
  {2021})}\BibitemShut {NoStop}%
\bibitem [{\citenamefont {Danielson}\ \emph {et~al.}(2021)\citenamefont
  {Danielson}, \citenamefont {Ghosh},\ and\ \citenamefont
  {Surko}}]{Danielson_2021}%
  \BibitemOpen
  \bibfield  {author} {\bibinfo {author} {\bibfnamefont {J.~R.}\ \bibnamefont
  {Danielson}}, \bibinfo {author} {\bibfnamefont {S.}~\bibnamefont {Ghosh}},\
  and\ \bibinfo {author} {\bibfnamefont {C.~M.}\ \bibnamefont {Surko}},\
  }\bibfield  {title} {\bibinfo {title} {Influence of geometry on positron
  binding to molecules},\ }\href {https://doi.org/10.1088/1361-6455/ac3e78}
  {\bibfield  {journal} {\bibinfo  {journal} {J. Phys. B}\ }\textbf {\bibinfo
  {volume} {54}},\ \bibinfo {pages} {225201} (\bibinfo {year}
  {2021})}\BibitemShut {NoStop}%
\bibitem [{\citenamefont {Ghosh}\ \emph {et~al.}(2022)\citenamefont {Ghosh},
  \citenamefont {Danielson},\ and\ \citenamefont {Surko}}]{Ghosh:2022}%
  \BibitemOpen
  \bibfield  {author} {\bibinfo {author} {\bibfnamefont {S.}~\bibnamefont
  {Ghosh}}, \bibinfo {author} {\bibfnamefont {J.~R.}\ \bibnamefont
  {Danielson}},\ and\ \bibinfo {author} {\bibfnamefont {C.~M.}\ \bibnamefont
  {Surko}},\ }\bibfield  {title} {\bibinfo {title} {Resonant annihilation and
  positron bound states in benzene},\ }\href
  {https://doi.org/10.1103/PhysRevLett.129.123401} {\bibfield  {journal}
  {\bibinfo  {journal} {Phys. Rev. Lett.}\ }\textbf {\bibinfo {volume} {129}},\
  \bibinfo {pages} {123401} (\bibinfo {year} {2022})}\BibitemShut {NoStop}%
\bibitem [{\citenamefont {Danielson}\ \emph {et~al.}(2022)\citenamefont
  {Danielson}, \citenamefont {Ghosh},\ and\ \citenamefont
  {Surko}}]{Danielson:2022}%
  \BibitemOpen
  \bibfield  {author} {\bibinfo {author} {\bibfnamefont {J.~R.}\ \bibnamefont
  {Danielson}}, \bibinfo {author} {\bibfnamefont {S.}~\bibnamefont {Ghosh}},\
  and\ \bibinfo {author} {\bibfnamefont {C.~M.}\ \bibnamefont {Surko}},\
  }\bibfield  {title} {\bibinfo {title} {Enhancement of positron binding energy
  in molecules containing $\ensuremath{\pi}$ bonds},\ }\href
  {https://doi.org/10.1103/PhysRevA.106.032811} {\bibfield  {journal} {\bibinfo
   {journal} {Phys. Rev. A}\ }\textbf {\bibinfo {volume} {106}},\ \bibinfo
  {pages} {032811} (\bibinfo {year} {2022})}\BibitemShut {NoStop}%
\bibitem [{\citenamefont {Gribakin}(2000)}]{PhysRevA.61.022720}%
  \BibitemOpen
  \bibfield  {author} {\bibinfo {author} {\bibfnamefont {G.~F.}\ \bibnamefont
  {Gribakin}},\ }\bibfield  {title} {\bibinfo {title} {Mechanisms of positron
  annihilation on molecules},\ }\href
  {https://doi.org/10.1103/PhysRevA.61.022720} {\bibfield  {journal} {\bibinfo
  {journal} {Phys. Rev. A}\ }\textbf {\bibinfo {volume} {61}},\ \bibinfo
  {pages} {022720} (\bibinfo {year} {2000})}\BibitemShut {NoStop}%
\bibitem [{\citenamefont {Gribakin}(2001)}]{Gribakin2001}%
  \BibitemOpen
  \bibfield  {author} {\bibinfo {author} {\bibfnamefont {G.~F.}\ \bibnamefont
  {Gribakin}},\ }\bibinfo {title} {Theory of positron annihilation on
  molecules},\ in\ \href {https://doi.org/10.1007/0-306-47613-4_22} {\emph
  {\bibinfo {booktitle} {New Directions in Antimatter Chemistry and
  Physics}}},\ \bibinfo {editor} {edited by\ \bibinfo {editor} {\bibfnamefont
  {C.~M.}\ \bibnamefont {Surko}}\ and\ \bibinfo {editor} {\bibfnamefont
  {F.~A.}\ \bibnamefont {Gianturco}}}\ (\bibinfo  {publisher} {Springer
  Netherlands},\ \bibinfo {address} {Dordrecht},\ \bibinfo {year} {2001})\ pp.\
  \bibinfo {pages} {413--435}\BibitemShut {NoStop}%
\bibitem [{\citenamefont {Tachikawa}\ \emph {et~al.}(2011)\citenamefont
  {Tachikawa}, \citenamefont {Kita},\ and\ \citenamefont
  {Buenker}}]{Tachikawa11}%
  \BibitemOpen
  \bibfield  {author} {\bibinfo {author} {\bibfnamefont {M.}~\bibnamefont
  {Tachikawa}}, \bibinfo {author} {\bibfnamefont {Y.}~\bibnamefont {Kita}},\
  and\ \bibinfo {author} {\bibfnamefont {R.~J.}\ \bibnamefont {Buenker}},\
  }\bibfield  {title} {\bibinfo {title} {Bound states of the positron with
  nitrile species with a configuration interaction multi-component molecular
  orbital approach},\ }\href {https://doi.org/10.1039/C0CP01650K} {\bibfield
  {journal} {\bibinfo  {journal} {Phys. Chem. Chem. Phys.}\ }\textbf {\bibinfo
  {volume} {13}},\ \bibinfo {pages} {2701} (\bibinfo {year}
  {2011})}\BibitemShut {NoStop}%
\bibitem [{\citenamefont {Tachikawa}(2014)}]{Tachikawa14}%
  \BibitemOpen
  \bibfield  {author} {\bibinfo {author} {\bibfnamefont {M.}~\bibnamefont
  {Tachikawa}},\ }\bibfield  {title} {\bibinfo {title} {Positron-attachment to
  acetonitrile, acetaldehyde, and acetone molecules: Vibrational enhancement of
  positron affinities with configuration interaction level of multi-component
  molecular orbital approach},\ }\href
  {https://doi.org/10.1088/1742-6596/488/1/012053} {\bibfield  {journal}
  {\bibinfo  {journal} {J. Phys.: Conf. Ser.}\ }\textbf {\bibinfo {volume}
  {488}},\ \bibinfo {pages} {012053} (\bibinfo {year} {2014})}\BibitemShut
  {NoStop}%
\bibitem [{\citenamefont {Romero}\ \emph
  {et~al.}(2014{\natexlab{a}})\citenamefont {Romero}, \citenamefont {Charry},
  \citenamefont {Flores-Moreno}, \citenamefont {Varella},\ and\ \citenamefont
  {Reyes}}]{APMO2014}%
  \BibitemOpen
  \bibfield  {author} {\bibinfo {author} {\bibfnamefont {J.}~\bibnamefont
  {Romero}}, \bibinfo {author} {\bibfnamefont {J.~A.}\ \bibnamefont {Charry}},
  \bibinfo {author} {\bibfnamefont {R.}~\bibnamefont {Flores-Moreno}}, \bibinfo
  {author} {\bibfnamefont {M.~T. d.~N.}\ \bibnamefont {Varella}},\ and\
  \bibinfo {author} {\bibfnamefont {A.}~\bibnamefont {Reyes}},\ }\bibfield
  {title} {\bibinfo {title} {Calculation of positron binding energies using the
  generalized any particle propagator theory},\ }\href
  {https://doi.org/10.1063/1.4895043} {\bibfield  {journal} {\bibinfo
  {journal} {J. Chem. Phys.}\ }\textbf {\bibinfo {volume} {141}},\ \bibinfo
  {pages} {114103} (\bibinfo {year} {2014}{\natexlab{a}})}\BibitemShut
  {NoStop}%
\bibitem [{\citenamefont {Swann}\ and\ \citenamefont
  {Gribakin}(2018)}]{Swann:2018}%
  \BibitemOpen
  \bibfield  {author} {\bibinfo {author} {\bibfnamefont {A.~R.}\ \bibnamefont
  {Swann}}\ and\ \bibinfo {author} {\bibfnamefont {G.~F.}\ \bibnamefont
  {Gribakin}},\ }\bibfield  {title} {\bibinfo {title} {Calculations of positron
  binding and annihilation in polyatomic molecules},\ }\href
  {https://doi.org/10.1063/1.5055724} {\bibfield  {journal} {\bibinfo
  {journal} {J. Chem. Phys.}\ }\textbf {\bibinfo {volume} {149}},\ \bibinfo
  {pages} {244305} (\bibinfo {year} {2018})}\BibitemShut {NoStop}%
\bibitem [{\citenamefont {Swann}\ and\ \citenamefont
  {Gribakin}(2019)}]{Swann:2019}%
  \BibitemOpen
  \bibfield  {author} {\bibinfo {author} {\bibfnamefont {A.~R.}\ \bibnamefont
  {Swann}}\ and\ \bibinfo {author} {\bibfnamefont {G.~F.}\ \bibnamefont
  {Gribakin}},\ }\bibfield  {title} {\bibinfo {title} {Positron binding and
  annihilation in alkane molecules},\ }\href
  {https://doi.org/10.1103/PhysRevLett.123.113402} {\bibfield  {journal}
  {\bibinfo  {journal} {Phys. Rev. Lett.}\ }\textbf {\bibinfo {volume} {123}},\
  \bibinfo {pages} {113402} (\bibinfo {year} {2019})}\BibitemShut {NoStop}%
\bibitem [{\citenamefont {Swann}\ and\ \citenamefont
  {Gribakin}(2020)}]{Swann:2020}%
  \BibitemOpen
  \bibfield  {author} {\bibinfo {author} {\bibfnamefont {A.~R.}\ \bibnamefont
  {Swann}}\ and\ \bibinfo {author} {\bibfnamefont {G.~F.}\ \bibnamefont
  {Gribakin}},\ }\bibfield  {title} {\bibinfo {title} {Model-potential
  calculations of positron binding, scattering, and annihilation for atoms and
  small molecules using a gaussian basis},\ }\href
  {https://doi.org/10.1103/PhysRevA.101.022702} {\bibfield  {journal} {\bibinfo
   {journal} {Phys. Rev. A}\ }\textbf {\bibinfo {volume} {101}},\ \bibinfo
  {pages} {022702} (\bibinfo {year} {2020})}\BibitemShut {NoStop}%
\bibitem [{\citenamefont {Suzuki}\ \emph {et~al.}(2020)\citenamefont {Suzuki},
  \citenamefont {Otomo}, \citenamefont {Iida}, \citenamefont {Sugiura},
  \citenamefont {Takayanagi},\ and\ \citenamefont {Tachikawa}}]{molbind_dft4}%
  \BibitemOpen
  \bibfield  {author} {\bibinfo {author} {\bibfnamefont {H.}~\bibnamefont
  {Suzuki}}, \bibinfo {author} {\bibfnamefont {T.}~\bibnamefont {Otomo}},
  \bibinfo {author} {\bibfnamefont {R.}~\bibnamefont {Iida}}, \bibinfo {author}
  {\bibfnamefont {Y.}~\bibnamefont {Sugiura}}, \bibinfo {author} {\bibfnamefont
  {T.}~\bibnamefont {Takayanagi}},\ and\ \bibinfo {author} {\bibfnamefont
  {M.}~\bibnamefont {Tachikawa}},\ }\bibfield  {title} {\bibinfo {title}
  {Positron binding in chloroethenes: Modeling positron-electron
  correlation-polarization potentials for molecular calculations},\ }\href
  {https://doi.org/10.1103/PhysRevA.102.052830} {\bibfield  {journal} {\bibinfo
   {journal} {Phys. Rev. A}\ }\textbf {\bibinfo {volume} {102}},\ \bibinfo
  {pages} {052830} (\bibinfo {year} {2020})}\BibitemShut {NoStop}%
\bibitem [{\citenamefont {Amaral}\ and\ \citenamefont
  {Mohallem}(2020)}]{MLbind}%
  \BibitemOpen
  \bibfield  {author} {\bibinfo {author} {\bibfnamefont {P.~H.~R.}\
  \bibnamefont {Amaral}}\ and\ \bibinfo {author} {\bibfnamefont {J.~R.}\
  \bibnamefont {Mohallem}},\ }\bibfield  {title} {\bibinfo {title}
  {Machine-learning predictions of positron binding to molecules},\ }\href
  {https://doi.org/10.1103/PhysRevA.102.052808} {\bibfield  {journal} {\bibinfo
   {journal} {Phys. Rev. A}\ }\textbf {\bibinfo {volume} {102}},\ \bibinfo
  {pages} {052808} (\bibinfo {year} {2020})}\BibitemShut {NoStop}%
\bibitem [{\citenamefont {Hofierka}\ \emph {et~al.}(2022)\citenamefont
  {Hofierka}, \citenamefont {Cunningham}, \citenamefont {Rawlins},
  \citenamefont {Patterson},\ and\ \citenamefont {Green}}]{Jaro22}%
  \BibitemOpen
  \bibfield  {author} {\bibinfo {author} {\bibfnamefont {J.}~\bibnamefont
  {Hofierka}}, \bibinfo {author} {\bibfnamefont {B.}~\bibnamefont
  {Cunningham}}, \bibinfo {author} {\bibfnamefont {C.~M.}\ \bibnamefont
  {Rawlins}}, \bibinfo {author} {\bibfnamefont {C.~H.}\ \bibnamefont
  {Patterson}},\ and\ \bibinfo {author} {\bibfnamefont {D.~G.}\ \bibnamefont
  {Green}},\ }\bibfield  {title} {\bibinfo {title} {Many-body theory of
  positron binding to polyatomic molecules},\ }\href
  {https://doi.org/https://doi.org/10.1038/s41586-022-04703-3} {\bibfield
  {journal} {\bibinfo  {journal} {Nature}\ }\textbf {\bibinfo {volume} {606}},\
  \bibinfo {pages} {688} (\bibinfo {year} {2022})}\BibitemShut {NoStop}%
\bibitem [{Note1()}]{Note1}%
  \BibitemOpen
  \bibinfo {note} {It also successfully described positron annihilation on
  small molecules \cite {Rawlins:2023}}\BibitemShut {NoStop}%
\bibitem [{Note2()}]{Note2}%
  \BibitemOpen
  \bibinfo {note} {Specifically, their model constructed the positron-molecule
  potential as a sum of positron-atom potentials, using a hybrid polarizability
  of an atom in a certain chemical environment. At long range the IPP takes the
  isotropic asymptotic form $-\alpha /2r^{4}$; while this is true for
  spherical-top molecules, the true asymptotic form of the polarization
  potential is generally anisotropic, and is given by $-1/2r^6 \DOTSI \sumop
  \slimits@ x_i x_j \alpha _{ij}$, where the $x_i$ are the Cartesian
  coordinates and $\alpha _{ij}$ are the Cartesian components of the
  polarizability tensor \cite {Swann:2018,molbind_dft4}}\BibitemShut {NoStop}%
\bibitem [{Note3()}]{Note3}%
  \BibitemOpen
  \bibinfo {note} {We use the fixed-nuclei approximation: compared to the
  correlations, vibrational effects have been found to have a relatively small
  effect on the binding energy \cite {Tachikawa14, PhysRevA.73.022705,
  Romero:2014, Buenker:2007, Buenker:2008,Jaro22}}\BibitemShut {NoStop}%
\bibitem [{Note4()}]{Note4}%
  \BibitemOpen
  \bibinfo {note} {Here the descriptor $(Z)$ refers to $(E)$/$(Z)$ isomerism,
  where $(Z)$ means that the highest priority groups, (i.e. the halogen atoms)
  are on the same side of the C-C double bond.}\BibitemShut {Stop}%
\bibitem [{\citenamefont {Dickhoff}\ and\ \citenamefont
  {Neck}(2008)}]{mbtexposed}%
  \BibitemOpen
  \bibfield  {author} {\bibinfo {author} {\bibfnamefont {W.~H.}\ \bibnamefont
  {Dickhoff}}\ and\ \bibinfo {author} {\bibfnamefont {D.~V.}\ \bibnamefont
  {Neck}},\ }\href@noop {} {\emph {\bibinfo {title} {Many-body {T}heory
  {E}xposed! - Propagator Description of Quantum Mechanics in Many-Body Systems
  - 2nd ed.}}}\ (\bibinfo  {publisher} {World Scientific, Singapore},\ \bibinfo
  {year} {2008})\BibitemShut {NoStop}%
\bibitem [{\citenamefont {Fetter}\ and\ \citenamefont
  {Walecka}(2003)}]{fetterwalecka}%
  \BibitemOpen
  \bibfield  {author} {\bibinfo {author} {\bibfnamefont {A.~L.}\ \bibnamefont
  {Fetter}}\ and\ \bibinfo {author} {\bibfnamefont {J.~D.}\ \bibnamefont
  {Walecka}},\ }\href@noop {} {\emph {\bibinfo {title} {Quantum theory of
  many-particle systems}}}\ (\bibinfo  {publisher} {Dover, New York},\ \bibinfo
  {year} {2003})\BibitemShut {NoStop}%
\bibitem [{\citenamefont {Bell}\ and\ \citenamefont
  {Squires}(1959)}]{PhysRevLett.3.96}%
  \BibitemOpen
  \bibfield  {author} {\bibinfo {author} {\bibfnamefont {J.~S.}\ \bibnamefont
  {Bell}}\ and\ \bibinfo {author} {\bibfnamefont {E.~J.}\ \bibnamefont
  {Squires}},\ }\bibfield  {title} {\bibinfo {title} {A formal optical model},\
  }\href {https://doi.org/10.1103/PhysRevLett.3.96} {\bibfield  {journal}
  {\bibinfo  {journal} {Phys. Rev. Lett.}\ }\textbf {\bibinfo {volume} {3}},\
  \bibinfo {pages} {96} (\bibinfo {year} {1959})}\BibitemShut {NoStop}%
\bibitem [{Note5()}]{Note5}%
  \BibitemOpen
  \bibinfo {note} {We use screened Coulomb interactions in the ladder diagrams
  and MO energies calculated in the random phase approximation. The
  $\varepsilon _b$ in this approximation are found to be within 4~meV of those
  calculated using bare Coulomb interactions and HF MO energies for chlorinated
  and brominated molecules (maximum relative error of $7.4\%$), and 0.1~meV for
  fluorinated molecules.}\BibitemShut {Stop}%
\bibitem [{\citenamefont {Kendall}\ \emph {et~al.}(1992)\citenamefont
  {Kendall}, \citenamefont {Dunning~Jr},\ and\ \citenamefont
  {Harrison}}]{Dunning}%
  \BibitemOpen
  \bibfield  {author} {\bibinfo {author} {\bibfnamefont {R.~A.}\ \bibnamefont
  {Kendall}}, \bibinfo {author} {\bibfnamefont {T.~H.}\ \bibnamefont
  {Dunning~Jr}},\ and\ \bibinfo {author} {\bibfnamefont {R.~J.}\ \bibnamefont
  {Harrison}},\ }\bibfield  {title} {\bibinfo {title} {Electron affinities of
  the first-row atoms revisited. {S}ystematic basis sets and wave functions},\
  }\href {https://doi.org/10.1063/1.462569} {\bibfield  {journal} {\bibinfo
  {journal} {J. Chem. Phys.}\ }\textbf {\bibinfo {volume} {96}},\ \bibinfo
  {pages} {6796} (\bibinfo {year} {1992})}\BibitemShut {NoStop}%
\bibitem [{Note6()}]{Note6}%
  \BibitemOpen
  \bibinfo {note} {The vPs contribution to the self energy requires
  diagonalization of dense matrices of size $(N_+N_-)^2$, where $N_+$ ($N_-$)
  is the number of virtual positron (electron) states used. For example, our
  C$_2$HCl$_3$ calculations have $N_+N_-=280800$, corresponding to $630$ GB of
  memory. With other memory costs considered, our approach regularly requires
  more than 1 TB of memory.}\BibitemShut {Stop}%
\bibitem [{\citenamefont {Young}\ and\ \citenamefont
  {Surko}(2008)}]{Young08_small}%
  \BibitemOpen
  \bibfield  {author} {\bibinfo {author} {\bibfnamefont {J.~A.}\ \bibnamefont
  {Young}}\ and\ \bibinfo {author} {\bibfnamefont {C.~M.}\ \bibnamefont
  {Surko}},\ }\bibfield  {title} {\bibinfo {title} {Feshbach-resonance-mediated
  positron annihilation in small molecules},\ }\href
  {https://doi.org/10.1103/PhysRevA.78.032702} {\bibfield  {journal} {\bibinfo
  {journal} {Phys. Rev. A}\ }\textbf {\bibinfo {volume} {78}},\ \bibinfo
  {pages} {032702} (\bibinfo {year} {2008})}\BibitemShut {NoStop}%
\bibitem [{\citenamefont {Gribakin}\ and\ \citenamefont {R.}(2006)}]{Gleb2006}%
  \BibitemOpen
  \bibfield  {author} {\bibinfo {author} {\bibfnamefont {G.~F.}\ \bibnamefont
  {Gribakin}}\ and\ \bibinfo {author} {\bibfnamefont {C.~M.~R.}\ \bibnamefont
  {Lee}},\ }\bibfield  {title} {\bibinfo {title} {Positron annihilation in
  molecules by capture into vibrational feshbach resonances of infrared-active
  modes},\ }\href {https://doi.org/10.1103/PhysRevLett.97.193201} {\bibfield
  {journal} {\bibinfo  {journal} {Phys. Rev. Lett.}\ }\textbf {\bibinfo
  {volume} {97}},\ \bibinfo {pages} {193201} (\bibinfo {year}
  {2006})}\BibitemShut {NoStop}%
\bibitem [{\citenamefont {Dzuba}\ and\ \citenamefont
  {Gribakin}(1994)}]{Dzuba:1994}%
  \BibitemOpen
  \bibfield  {author} {\bibinfo {author} {\bibfnamefont {V.~A.}\ \bibnamefont
  {Dzuba}}\ and\ \bibinfo {author} {\bibfnamefont {G.~F.}\ \bibnamefont
  {Gribakin}},\ }\bibfield  {title} {\bibinfo {title} {Correlation-potential
  method for negative ions and electron scattering},\ }\href
  {https://doi.org/10.1103/PhysRevA.49.2483} {\bibfield  {journal} {\bibinfo
  {journal} {Phys. Rev. A}\ }\textbf {\bibinfo {volume} {49}},\ \bibinfo
  {pages} {2483} (\bibinfo {year} {1994})}\BibitemShut {NoStop}%
\bibitem [{\citenamefont {Gribakin}\ and\ \citenamefont
  {Swann}(2015)}]{Gribakin2015}%
  \BibitemOpen
  \bibfield  {author} {\bibinfo {author} {\bibfnamefont {G.~F.}\ \bibnamefont
  {Gribakin}}\ and\ \bibinfo {author} {\bibfnamefont {A.~R.}\ \bibnamefont
  {Swann}},\ }\bibfield  {title} {\bibinfo {title} {Effect of dipole
  polarizability on positron binding by strongly polar molecules},\ }\href
  {https://doi.org/10.1088/0953-4075/48/21/215101} {\bibfield  {journal}
  {\bibinfo  {journal} {J. Phys. B}\ }\textbf {\bibinfo {volume} {48}},\
  \bibinfo {pages} {215101} (\bibinfo {year} {2015})}\BibitemShut {NoStop}%
\bibitem [{\citenamefont {Green}\ and\ \citenamefont
  {Gribakin}(2015)}]{DGG:2015:core}%
  \BibitemOpen
  \bibfield  {author} {\bibinfo {author} {\bibfnamefont {D.~G.}\ \bibnamefont
  {Green}}\ and\ \bibinfo {author} {\bibfnamefont {G.~F.}\ \bibnamefont
  {Gribakin}},\ }\bibfield  {title} {\bibinfo {title} {$\ensuremath{\gamma}$
  spectra and enhancement factors for positron annihilation with core
  electrons},\ }\href {https://doi.org/10.1103/PhysRevLett.114.093201}
  {\bibfield  {journal} {\bibinfo  {journal} {Phys. Rev. Lett.}\ }\textbf
  {\bibinfo {volume} {114}},\ \bibinfo {pages} {093201} (\bibinfo {year}
  {2015})}\BibitemShut {NoStop}%
\bibitem [{\citenamefont {Green}\ and\ \citenamefont
  {Gribakin}(2018)}]{DGG:2017:ef}%
  \BibitemOpen
  \bibfield  {author} {\bibinfo {author} {\bibfnamefont {D.~G.}\ \bibnamefont
  {Green}}\ and\ \bibinfo {author} {\bibfnamefont {G.~F.}\ \bibnamefont
  {Gribakin}},\ }\bibfield  {title} {\bibinfo {title} {Enhancement factors for
  positron annihilation on valence and core orbitals of noble-gas atoms},\
  }\href {https://doi.org/"10.1007/978-3-319-74582-4_14"} {\bibfield  {journal}
  {\bibinfo  {journal} {Concepts, Methods and Applications of Quantum Systems
  in Chemistry and Physics, Prog. Theor. Chem. and Phys.}\ }\textbf {\bibinfo
  {volume} {31}},\ \bibinfo {pages} {243} (\bibinfo {year} {2018})}\BibitemShut
  {NoStop}%
\bibitem [{\citenamefont {Rawlins}\ \emph {et~al.}(2023)\citenamefont
  {Rawlins}, \citenamefont {Hofierka}, \citenamefont {Cunningham},
  \citenamefont {Patterson},\ and\ \citenamefont {Green}}]{Rawlins:2023}%
  \BibitemOpen
  \bibfield  {author} {\bibinfo {author} {\bibfnamefont {C.~M.}\ \bibnamefont
  {Rawlins}}, \bibinfo {author} {\bibfnamefont {J.}~\bibnamefont {Hofierka}},
  \bibinfo {author} {\bibfnamefont {B.}~\bibnamefont {Cunningham}}, \bibinfo
  {author} {\bibfnamefont {C.~H.}\ \bibnamefont {Patterson}},\ and\ \bibinfo
  {author} {\bibfnamefont {D.~G.}\ \bibnamefont {Green}},\ }\bibfield  {title}
  {\bibinfo {title} {Many-body theory calculations of positron scattering and
  annihilation in {${\mathrm{H}}_{2}$}, {${\mathrm{N}}_{2}$}, and
  {${\mathrm{CH}}_{4}$}},\ }\href
  {https://doi.org/10.1103/PhysRevLett.130.263001} {\bibfield  {journal}
  {\bibinfo  {journal} {Phys. Rev. Lett.}\ }\textbf {\bibinfo {volume} {130}},\
  \bibinfo {pages} {263001} (\bibinfo {year} {2023})}\BibitemShut {NoStop}%
\bibitem [{\citenamefont {Gianturco}\ \emph {et~al.}(2006)\citenamefont
  {Gianturco}, \citenamefont {Franz}, \citenamefont {Buenker}, \citenamefont
  {Liebermann}, \citenamefont {Pichl}, \citenamefont {Rost}, \citenamefont
  {Tachikawa},\ and\ \citenamefont {Kimura}}]{PhysRevA.73.022705}%
  \BibitemOpen
  \bibfield  {author} {\bibinfo {author} {\bibfnamefont {F.~A.}\ \bibnamefont
  {Gianturco}}, \bibinfo {author} {\bibfnamefont {J.}~\bibnamefont {Franz}},
  \bibinfo {author} {\bibfnamefont {R.~J.}\ \bibnamefont {Buenker}}, \bibinfo
  {author} {\bibfnamefont {H.-P.}\ \bibnamefont {Liebermann}}, \bibinfo
  {author} {\bibfnamefont {L.~c.~v.}\ \bibnamefont {Pichl}}, \bibinfo {author}
  {\bibfnamefont {J.-M.}\ \bibnamefont {Rost}}, \bibinfo {author}
  {\bibfnamefont {M.}~\bibnamefont {Tachikawa}},\ and\ \bibinfo {author}
  {\bibfnamefont {M.}~\bibnamefont {Kimura}},\ }\bibfield  {title} {\bibinfo
  {title} {Positron binding to alkali-metal hydrides: The role of molecular
  vibrations},\ }\href {https://doi.org/10.1103/PhysRevA.73.022705} {\bibfield
  {journal} {\bibinfo  {journal} {Phys. Rev. A}\ }\textbf {\bibinfo {volume}
  {73}},\ \bibinfo {pages} {022705} (\bibinfo {year} {2006})}\BibitemShut
  {NoStop}%
\bibitem [{\citenamefont {Romero}\ \emph
  {et~al.}(2014{\natexlab{b}})\citenamefont {Romero}, \citenamefont {Charry},
  \citenamefont {Flores-Moreno}, \citenamefont {Varella},\ and\ \citenamefont
  {Reyes}}]{Romero:2014}%
  \BibitemOpen
  \bibfield  {author} {\bibinfo {author} {\bibfnamefont {J.}~\bibnamefont
  {Romero}}, \bibinfo {author} {\bibfnamefont {J.~A.}\ \bibnamefont {Charry}},
  \bibinfo {author} {\bibfnamefont {R.}~\bibnamefont {Flores-Moreno}}, \bibinfo
  {author} {\bibfnamefont {M.}~\bibnamefont {Varella}},\ and\ \bibinfo {author}
  {\bibfnamefont {A.}~\bibnamefont {Reyes}},\ }\bibfield  {title} {\bibinfo
  {title} {Calculation of positron binding energies using the generalized any
  particle propagator theory},\ }\href {https://doi.org/10.1063/1.4895043}
  {\bibfield  {journal} {\bibinfo  {journal} {J. Chem. Phys.}\ }\textbf
  {\bibinfo {volume} {141}},\ \bibinfo {pages} {114103} (\bibinfo {year}
  {2014}{\natexlab{b}})}\BibitemShut {NoStop}%
\bibitem [{\citenamefont {Buenker}\ and\ \citenamefont
  {Libermann}(2007)}]{Buenker:2007}%
  \BibitemOpen
  \bibfield  {author} {\bibinfo {author} {\bibfnamefont {R.~J.}\ \bibnamefont
  {Buenker}}\ and\ \bibinfo {author} {\bibfnamefont {H.}~\bibnamefont
  {Libermann}},\ }\bibfield  {title} {\bibinfo {title} {Role of the electric
  dipole moment in positron binding to the ground and excited states of the
  {B}e{O} molecule},\ }\href {https://doi.org/10.1063/1.2711203} {\bibfield
  {journal} {\bibinfo  {journal} {J. Chem. Phys.}\ }\textbf {\bibinfo {volume}
  {126}},\ \bibinfo {pages} {104305} (\bibinfo {year} {2007})}\BibitemShut
  {NoStop}%
\bibitem [{\citenamefont {Buenker}\ and\ \citenamefont
  {Liebermann}(2008)}]{Buenker:2008}%
  \BibitemOpen
  \bibfield  {author} {\bibinfo {author} {\bibfnamefont {R.~J.}\ \bibnamefont
  {Buenker}}\ and\ \bibinfo {author} {\bibfnamefont {H.-P.}\ \bibnamefont
  {Liebermann}},\ }\bibfield  {title} {\bibinfo {title} {Configuration
  interaction calculations of positron binding to molecular oxides and hydrides
  and its effect on spectroscopic constants},\ }\href
  {https://doi.org/https://doi.org/10.1016/j.nimb.2007.12.029} {\bibfield
  {journal} {\bibinfo  {journal} {Nucl. Instrum. and Meth. B}\ }\textbf
  {\bibinfo {volume} {266}},\ \bibinfo {pages} {483} (\bibinfo {year}
  {2008})}\BibitemShut {NoStop}%
\end{thebibliography}



\foreach \x in {1,2,3,4}
{%
\clearpage


\section*{Supplementary material}

# Many-body Theory Calculations of Positron Binding to Halogenated Hydrocarbons - Supplementary Material

J. P. Cassidy,[1, *] J. Hofierka,[1] B. Cunningham,[1] C. M. Rawlins,[1] C. H. Patterson,[2] and D. G. Green[1, †]

[1]*School of Mathematics and Physics, Queen's University Belfast, University Road, Belfast BT7 1NN, United Kingdom*
[2]*School of Physics, Trinity College Dublin, Dublin 2, Ireland*
(Dated: December 20, 2023)

**Further details of the basis sets and placement and effect of ghosts on $\varepsilon_b$ in the calculations**.—We expand the electron and positron wave functions in Gaussian basis sets, using aug-cc-pVXZ bases (X=T,Q) [1] on atomic centres as well as additional hydrogen aug-cc-pVXZ bases on "ghost" centres 1Å from the molecule to resolve regions of maximum positron density. We also use diffuse even-tempered positron bases of the form $10s9p8d7f6g$, with exponents $\zeta_0 \times \beta^{k-1}$ ($\zeta_0 = 0.00001$– $0.006$ and $\beta = 2$– $3$), ensuring the positron is described well at large distances $r \sim 1/\kappa$, where $\kappa = \sqrt{2\varepsilon_b}$. Regarding the arrangement of ghost basis centres ('ghost atoms') used, we can include two distinct types of ghost atom, each of which holding a distinct basis set. We refer to these as G1 (ghost 1) and G2 (ghost 2) atoms. Ghosts carry both positron and electron basis functions. For all of the molecules considered, we placed 5 G1 ghosts around each halogen atom in the molecule in the shape of a square-pyramidal cap, with each 1Å away from the halogen, and the halogen lying at the centre of the square base of the pyramid (see configuration 4 in Table I). We also included a single G2 ghost holding the even tempered positron basis to improve the description of the long-range positron-molecule interactions. The calculation for *trans*-1,2-dichloroethylene, for which the Cl atoms are on opposite sides of the molecule, used this pyramidal arrangement of ghosts, but four extra G1 ghosts were also included, each 1Å above and below each carbon in the molecule. The only exceptions to the addition of a G2 even tempered ghost were $CCl_4$ and $CF_4$; instead their even tempered bases for the positron were placed on the central carbon atom in attempt to save computational resources (to compensate for the use of 20 G1 ghosts). The effect of the number and placement of G1 ghost basis centres on the positron binding energy calculation was investigated. We performed calculations for several G1 ghost configurations for chloromethane $CH_3Cl$, ranging from including 2–9 G1 ghosts. Every configuration also contained an even-tempered G2 ghost placed in between the C and Cl atoms. The results are shown in Supplemental Table I. In the 5 ghost pyramid arrangement we see that placing the ghosts 1Å away from the halogen gives a higher binding energy compared to placing the ghosts 1.5Å away, suggesting 1Å is the more suitable distance. Convergence with respect to the addition of ghosts was achieved by the time 7 additional ghosts were used, with $\varepsilon_b = 27$ meV. For larger molecules with several carbon and chlorine atoms, it would not have been computationally feasible to use 7 ghosts per halogen, hence we adhered to using 5 throughout. For molecules with 2 or more halogens this would naturally lead to there being 10 or more additional G1 ghosts in the calculation anyway, so convergence with respect to ghosts should not suffer drastically. For $CHCl_3$, we tried additional BSE+Γ+Λ calculations where we manually adjusted the exponents of the G1 positron basis functions so that the FWHMs of the Gaussians were scaled by factors of 0.5, 2, and 4 to investigate the sensitivity of the results without the additional computational cost of added basis functions. These yielded binding energies of 26, 25, and 24 meV, little to no improvement on the 25 meV reported in the main text.

**Renormalization constants**.—The positron Dyson wavefunction is a quasiparticle wave function, that is, the overlap of the wave function of the *N*-electron ground state molecule with the fully-correlated wave function of the positron plus *N*-electron molecule system. Reflecting its many-body nature, it satisfies

$$\int |\psi_\varepsilon(\mathbf{r})|^2 d\mathbf{r} = \left(1 - \partial\varepsilon/\partial E|_{\varepsilon_b}\right)^{-1} \equiv a < 1. \quad (1)$$

where $\varepsilon$ is the energy eigenvalue of the solution to the Dyson equation, $E$ the energy at which the self energy $\Sigma_E$ is calculated, and $a$ is the renormalization constant whose value quantifies the degree to which the positron-molecule bound state is a single-particle state, with smaller values of $a$ signifying more strongly-correlated states. The renormalization factors for the molecules in the main text are given in Table II.

**Effects of halogenation**.—Supplemental Figure 1 shows the Hartree-Fock molecular orbital energies of the halogenated hydrocarbons for which we performed *ab initio* calculations. Comparing the brominated and chlorinated molecules to fluorinated molecules, we see that in general the states in the fluorinated molecules are at higher energies, indicating more tightly bound electrons. For those molecules where the molecular ionization energies (i.e., the HOMO energy) are similar, there is a larger gap between the HOMO and the next highest occupied orbital in the fluorinated molecule than in the brominated or chlorinated molecule. We see the same trends in the bromochloro and chlorofluoro intermediate molecules. Overall this inhibits the positron's ability to polarize the electron cloud, and makes virtual Ps formation more difficult, as tightly bound electrons are less easily

TABLE I. Calculated positron binding energies (in meV) for chloromethane at the $\Sigma^{GW+\Gamma+\Lambda}$ level of many-body theory compared with experiment using several configurations of G1 ghost atoms (pink atoms in diagrams).

| | Chloromethane CH$_3$Cl | | |
|---|---|---|---|
| Ghost configuration | | $\Sigma^{GW+\Gamma+\Lambda}$[a] | Exp. [2] |
| 2×G1 collinear to C-Cl | 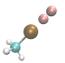 | 15 | 26 ± 6 |
| 2×G1 perp. to C-Cl | 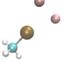 | 16 | 26 ± 6 |
| 3×G1 | 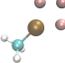 | 20 | 26 ± 6 |
| 5×G1 1Å from Cl | 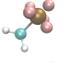 | 25 | 26 ± 6 |
| 5×G1 1.5Å from Cl | 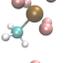 | 24 | 26 ± 6 |
| 7×G1 square & triangular plane | 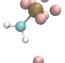 | 27 | 26 ± 6 |
| 8×G1 two square planes | 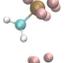 | 27 | 26 ± 6 |
| 9×G1 two square planes plus apex | 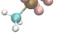 | 27 | 26 ± 6 |

[a] Using screened Coulomb interactions and $GW$@RPA energies in the ladder diagrams for $\Sigma^\Gamma$ and $\Sigma^\Lambda$.

TABLE II. Calculated renormalization factors, $a$, for the positron Dyson wave functions at the $\Sigma^{GW+\Gamma+\Lambda}$ level of many-body theory[a].

| Molecule | Formula | $a$ |
|---|---|---|
| Chloromethane | CH$_3$Cl | 0.986 |
| Dichloromethane | CH$_2$Cl$_2$ | 0.986 |
| Trichloromethane | CHCl$_3$ | 0.984 |
| Tetrachloromethane | CCl$_4$ | 0.980 |
| Vinyl chloride | C$_2$H$_3$Cl | 0.982 |
| Vinylidene chloride | C$_2$H$_2$Cl$_2$ | 0.979 |
| cis-1,2-dichloroethylene | C$_2$H$_2$Cl$_2$ | 0.974 |
| trans-1,2-dichloroethylene | C$_2$H$_2$Cl$_2$ | 0.988 |
| Trichloroethylene | C$_2$HCl$_3$ | 0.979 |
| 1-chloro-1-fluoroethylene | C$_2$H$_2$ClF | 0.994 |
| (Z)-chlorofluoroethylene | C$_2$H$_2$ClF | 0.986 |
| Fluoromethane | CH$_3$F | 0.999 |
| Difluoromethane | CH$_2$F$_2$ | 0.999 |
| Vinyl fluoride | C$_2$H$_3$F | 0.999 |
| cis-1,2-difluoroethylene | C$_2$H$_2$F$_2$ | 0.997 |
| Bromomethane | CH$_3$Br | 0.976 |
| cis-1,2-dibromoethylene | C$_2$H$_2$Br$_2$ | 0.942 |
| (Z)-bromochloroethylene | C$_2$H$_2$BrCl | 0.961 |

[a] Using screened Coulomb interactions and $GW$@RPA energies in the ladder diagrams for $\Sigma^\Gamma$ and $\Sigma^\Lambda$.

perturbed by the positron and will not tunnel to the positron as readily.

$Z_{\text{eff}}$ **spectrum of CH$_3$Br.**—For vibrational-Feshbach resonant annihilation, the positron-momentum-dependent annihilation spectrum can be estimated by the Gribakin-Lee model [3], viz.,

$$Z_{\text{eff}}^{(\text{res})}(\varepsilon) = 2\pi^2 \delta_{ep} \sum_\nu \frac{g_\nu \Gamma_\nu^e}{k_\nu \Gamma_\nu} \Delta(\varepsilon - \varepsilon_\nu), \quad (2)$$

where $\varepsilon$ is the incident positron's energy (with momentum $k = \sqrt{2\varepsilon}$), $\delta_{ep}$ is the contact density, $\nu$ is a vibrational mode of the

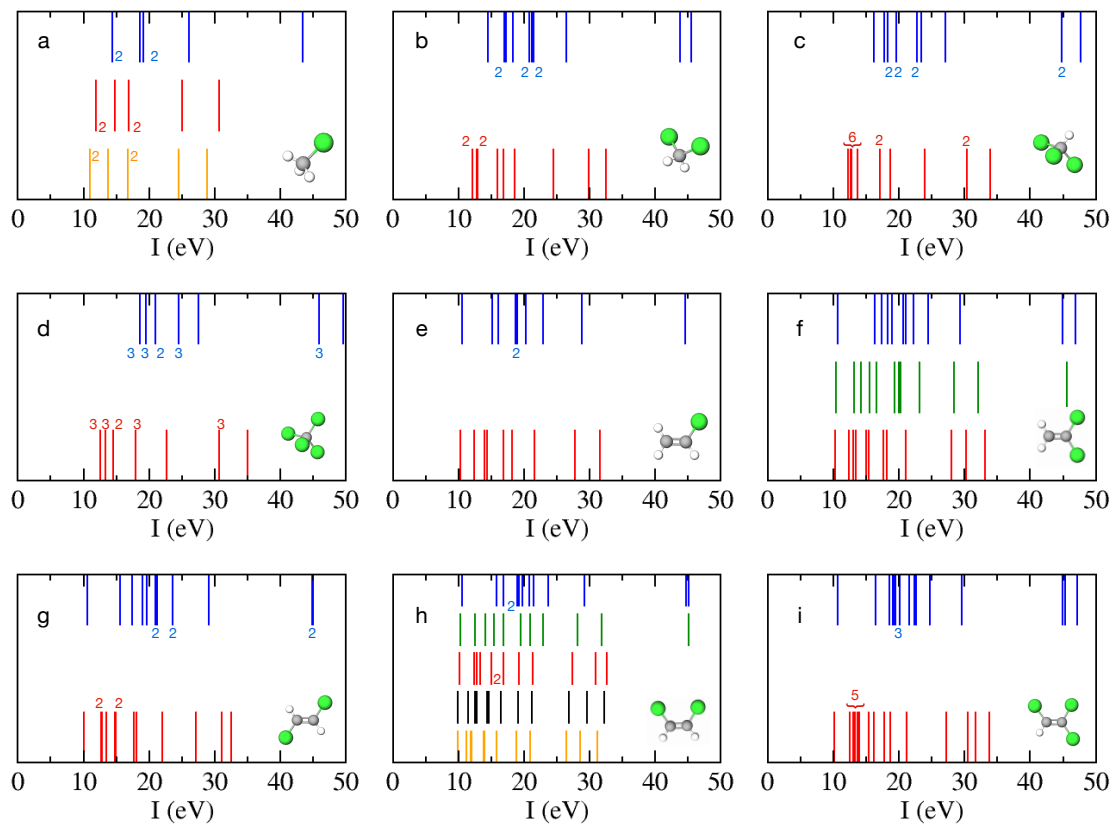

Supplemental Fig. 1. Hartree-Fock molecular orbital energies of the halogenated hydrocarbons for which we performed *ab initio* calculations. Orange lines, brominated molecules; black lines bromo-chloro intermediate molecules; red lines, chlorinated molecules; green lines, chloro-fluoro intermediate molecules; blue lines, fluorinated molecules: a) bromomethane, chloromethane, fluoromethane; b) dichloromethane, difluoromethane; c) trichloromethane, trifluoromethane; d) tetrachloromethane, tetrafluoromethane; e) vinyl chloride, vinyl fluoride; f) vinylidene chloride, vinylidene fluoride, 1-chloro-1-fluoroethylene; g) *trans*-1,2-dichloroethylene, *trans*-1,2-difluoroethylene; h) *cis*-1,2-dibromoethylene, (Z)-bromochloroethylene, *cis*-1,2-dichloroethylene, (Z)-chlorofluoroethylene, *cis*-1,2-difluoroethylene; i) trichloroethylene, trifluoroethylene. Green atoms represent the positions of the halogen(s) in each structure. Coloured numerals are used to indicate closely lying energy states (indistinguishable on the plot) and energy degeneracies.

molecule with degeneracy $g_\nu$, $k_\nu$ is related to the energy of the mode as $k_\nu^2/2 = \omega_\nu - |\varepsilon_b|$, $\Gamma_\nu$ and $\Gamma_\nu^e$ are the total and elastic resonance widths respectively (and their ratio is close to unity), and $\Delta(E)$ is related to the energy distribution of the positrons in the experimental beam. Supplemental Figure 2 shows the calculated $Z_{\text{eff}}(\varepsilon)$ spectrum using our calculated values of $\varepsilon_b = 56$ meV and $\delta_{ep} = 1.139 \times 10^{-2}$ a.u. as the free parameters of the model. For comparison we also show the experimental spectrum, and the original result using the Gribakin-Lee model that used $\varepsilon_b = 40$ meV and assumed that the contact density was proportional to the square root of the binding energy as $\delta_{ep} = (0.66/2\pi)\sqrt{2\varepsilon_b}$. As shown in Supplemental Figure 2, our calculated $Z_{\text{eff}}$ spectrum is downshifted and slightly enhanced relative to the original Gribakin-Lee model. It was recently reported that some of the earlier measurements of positron binding energies may have contained systematic errors [2]– a new measurement for $CH_3Br$ would therefore be of interest.

* jcassidy18@qub.ac.uk
† d.green@qub.ac.uk

[1] R. A. Kendall, T. H. Dunning Jr, and R. J. Harrison, Electron affinities of the first-row atoms revisited. Systematic basis sets and wave functions, J. Chem. Phys. **96**, 6796 (1992).

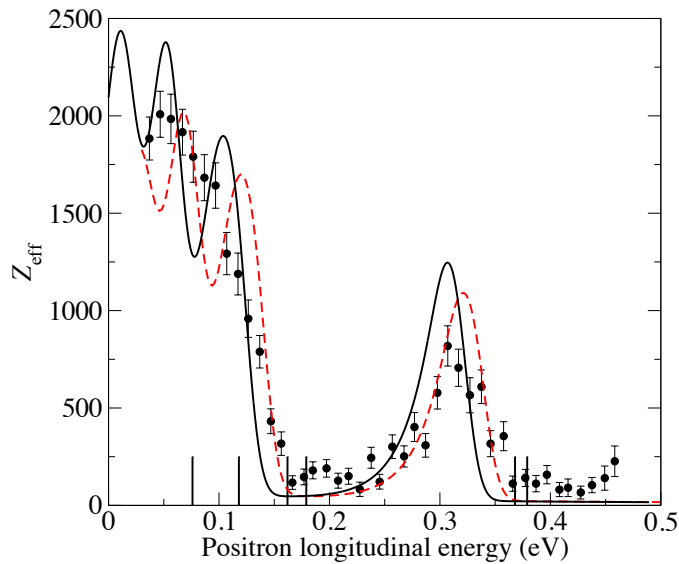

Supplemental Fig. 2. $Z_{\text{eff}}$ annihilation spectrum of CH$_3$Br as a function of the incident positron beam's longitudinal energy. Black symbols are experiment [4]; red dashed curve is the Gribakin-Lee model using an experimental binding energy of $\varepsilon_b$ = 40 meV [3]; Black solid line the Gribakin-Lee model, but using the presently calculated binding energy $\varepsilon_b$ = 56 meV and enhanced and renormalized contact density $\delta_{ep} = 5.940 \times 10^{-3}$ a.u. Vertical bars show the energies of molecular fundamental vibrational modes. The peak near 0.3 eV corresponds to the C-H stretch mode.

[2] A. R. Swann, G. F. Gribakin, J. R. Danielson, S. Ghosh, M. R. Natisin, and C. M. Surko, Effect of chlorination on positron binding to hydrocarbons: experiment and theory, Phys. Rev. A **104**, 012813 (2021).
[3] G. F. Gribakin and L. C. M. R., Positron annihilation in molecules by capture into vibrational feshbach resonances of infrared-active modes, Phys. Rev. Lett. **97**, 193201 (2006).
[4] L. D. Barnes, J. A. Young, and C. M. Surko, Energy-resolved positron annihilation rates for molecules, Phys. Rev. A **74**, 012706 (2006).


}

\end{document}